\documentclass[man,nonatbib,hidelinks,12pt]{elsarticle}

\AtBeginDocument{\RequirePackage{xurl}}
\usepackage[breaklinks]{hyperref}
\usepackage{booktabs}

\usepackage{graphicx}

\usepackage{xcolor}

\usepackage{csquotes}

\usepackage[american]{babel}

\usepackage[square,numbers,sort]{natbib}
\journal{Natural Language Processing Journal}

\begin{document}

\begin{frontmatter}

\title{Cutting through the noise to motivate people: A comprehensive analysis of COVID-19 social media posts de/motivating vaccination}

\begin{abstract}
    The COVID-19 pandemic exposed significant weaknesses in the healthcare information system. The overwhelming volume of misinformation on social media and other socioeconomic factors created extraordinary challenges to motivate people to take proper precautions and get vaccinated. In this context, our work explored a novel direction by analyzing an extensive dataset collected over two years, identifying the topics de/motivating the public about COVID-19 vaccination. We analyzed these topics based on time, geographic location, and political orientation. We noticed that while the motivating topics remain the same over time and geographic location, the demotivating topics rapidly. We also identified that intrinsic motivation, rather than external mandate, is more advantageous to inspire the public. This study addresses scientific communication and public motivation in social media. It can help public health officials, policymakers, and social media platforms develop more effective messaging strategies to cut through the noise of misinformation and educate the public about scientific findings.
\end{abstract}


\author[niu]{Ashiqur Rahman\corref{cor1}}
\ead{ashiqur.r@niu.edu}

\author[usc]{Ehsan Mohammadi}
\ead{ehsan2@sc.edu}

\author[niu]{Hamed Alhoori}
\ead{alhoori@niu.edu}

\cortext[cor1]{Corresponding author}

\affiliation[niu]{organization={Department of Computer Science, Northern Illinois University},
            addressline={100 Normal Road},
            city={DeKalb},
            postcode={60115},
            state={IL},
            country={USA}}

\affiliation[usc]{organization={School of Information Science, University of South Carolina},
            addressline={800 Sumter Street},
            city={Columbia},
            postcode={29208},
            state={SC},
            country={USA}}

\end{frontmatter}

\section{Introduction} \label{sec:introduction}
Social media plays a vital role in modern life by availing communication, information dissemination, and steering social conversations \citep{Ye2020-ht}. These features are more impactful during a state of crisis \citep{Palen-2008-ds}. We have seen this during major social events like mass shootings, natural disasters, national elections, and even in anti-vaccination campaigns \citep{Badawy2018-xc, Broniatowski2018-xv, Housholder2015-wb, Palen-2008-ds}.

The Coronavirus pandemic created significant dependence on social media. While the social web was essential for disseminating healthcare information, important announcements, and educating the public, misinformation has also spread with little oversight. Studies suggest that exposure to information on social media greatly influenced preventative behavior during the COVID-19 pandemic \citep{Yoo2023-gu, Kim2020-cg}. Hence, the excessive dissemination of misinformation on social media was a significant concern. As soon as the coronavirus emerged, racism, rumor, and fear-mongering started spreading like wildfire on different platforms \citep{Depoux2020-ay}. Although there were projects such as Poynter \citep{Poynter-Institute2020-ol} and EUvsDisinfo \citep{East-StratCom-Task-Force2015-xy}, actively monitoring and debunking false news, misinformation on social media was widely available. The World Health Organization (WHO) partnered with major tech giants such as Facebook, Google, LinkedIn, Microsoft, Reddit, and Twitter to fight against misinformation \citep{Tasnim2020-sn}. However, misinformation was still widely available on these platforms. WHO director-general called it the fight against `trolls and conspiracy theories.' 

While vaccination is one of the most effective tools for preventing diseases and keeping communities safe \citep{Office-of-Infectious-Disease2021-pz, Gerberding2021-ks}, it needed a significant percentage of the population to be vaccinated to achieve herd immunity \citep{Who2020-fu} and leave the pandemic behind us. For comparison, measles requires 95\% vaccination, and polio requires 80\% vaccination coverage to achieve this collective immunity \citep{Who2020-fu}. Research suggests that although it might be unattainable to achieve herd immunity in the traditional sense against COVID-19 \citep{Morens2022-fx}, a high vaccination rate reduces the disease's effect and makes it more manageable \citep{DSouza2021-fo} and returns our life to normalcy. 

Once different vaccines were accessible to the public around the world \citep{Who2020-ck}, motivating enough people to get vaccinated quickly became a challenge. Different factors caused vaccine hesitancy, including but not limited to public trust in the development and approval of vaccines, economic disparity, education, and ethnicity \citep{Daly2021-nl, Kricorian2021-bf, Murphy2021-yp, Truong2022-qs, Guidry2021-nl}. Misinformation on social media played a vital role in emboldening the misconception about vaccination. For instance, a survey in the UK revealed that people who relied on social web platforms to acquire information were more reluctant to receive vaccines \citep{Nielsen2020-rw}. Similarly, another study confirmed the connection between the cluster of the unconvinced public on Facebook and the networks of anti-vaxxers \citep{Johnson2020-ev}. The understanding of fairness and transparency of the social media platform also impacts the vaccination decision \cite{Villacis-Calderon2023-bm}. Research also suggests that COVID-19 was highly politicized in the mainstream news media \citep{Abbas2020-bg, Hart2020-pu}, contributing to demotivation and distrust of the vaccination \citep{Thelwall2021-fg}. The study by \citet{Liu2023-yw} shows that the public is less likely to follow the directives from authority when the message is politically polarizing. Delays in achieving the vaccination target had economic impacts as well. A study found that the total monetary harm from ``non-vaccination" in the United States was between 50 to 300 million dollars per day \citep{Bruns2021-tf}. Studies also suggested that disproving misinformation was insufficient, and rebuilding public trust in government institutions and scientific processes was essential \citep{Guidry2021-nl}. 

With the advancement of Large Language Models (LLMs) and their accessibility, which can quickly generate believable but fictional scientific texts without proper references or scientific knowledge, the risk of misguiding the public has become even more significant. This makes it essential to find ways to reach the public with factual information, breaking through the clutter of misinformation and disinformation. \citep{Anonymous2023-nq, Zhou2023-it}

Therefore, it is crucial to study, identify, and address the factors that demotivate the public and increase hesitancy and distrust towards vaccination. The knowledge gathered from this study can be helpful to policymakers, healthcare workers, and social media platforms to improve the handling of misinformation and alleviate doubts in the community during emergencies.

In this study, we especially focused on the de/motivation of getting vaccines during the COVID-19 pandemic and social media's role in it. We aim to explore this issue on Twitter with the following research questions:

\begin{enumerate}
    \item What were the most popular topics on Twitter that were de/motivating people about the COVID-19 vaccine?  \label{rq:rq1}
    \item Which topics are influencing the public stance towards COVID-19 vaccination? \label{rq:rq2}
    \item Do the motivating and demotivating topics about the COVID-19 vaccine on Twitter change based on time, geographic location, or political landscape within the US? \label{rq:rq3}
\end{enumerate}

\subsection{Contributions} \label{sec:contributions}

Throughout the study, we answered the questions above and delivered the following data and machine-learning models. We shared the models and data generated from this work with the research community.
\begin{itemize}
    \item A labeled Twitter dataset spanning from January 2020 to December 2021 containing location, motivating status, vaccination stance, and topic label.
    \item An analysis of the COVID-19 vaccination topic distribution over time, US states, and political orientation of different states.
    \item A machine-learning model to classify tweets as motivating or demotivating about the COVID-19 vaccine.
    \item A machine-learning model to identify the COVID-19 vaccination stance of a user based on their tweets.
    \item Topic models to identify COVID-19 vaccination-related motivating and demotivating topics on Twitter.
\end{itemize}

In the next section, we discussed related works in this field, followed by a detailed explanation of the data collection process. Then, we discussed the methods for de/motivation classification of the tweets and topic extraction, followed by models for stance detection. We ended the paper with a discussion about our work, limitations, ethical concerns, and conclusion. We also included all the datasets in \ref{sec:datasets}, models in \ref{sec:machinelearningmodels}, and environmental parameters in \ref{sec:parametersforreproducability} for reproducibility purposes.

\section{Related Works} \label{sec:relatedworks}
\subsection{Misinformation and Vaccine} \label{sec:misinformationandvaccine}
For better or worse, social media has become an increasingly popular method for everyday people to obtain information on various topics like scientific findings, current events, news, political occurrences, and many more. Social media can be an effective way for individuals to stay connected with the outside world and each other. Still, any user can post whatever content they desire, regardless of the validity of the information that the post contains. Thus, a paradox emerges in which everyday people have access to more information at their fingertips than ever before and an increased propensity for exposure to misinformation. This creates a chaotic information landscape characterized by a general inability of people to distinguish between fact and fiction in the pieces of information they encounter. This whirlwind of disseminated information and misinformation dramatically impacts the overall public perception involving specific issues. Many researchers have attempted to analyze the relationship between social media trends and public opinion regarding public health issues, like vaccination and immunization programs \citep{Karafillakis2021-vy, Suarez-Lledo2021-uf, Mota2021-lk, Becker2016-uw, Blankenship2018-fl, Basch2017-ac}.

Several researchers have conducted analyses of Twitter content to determine the general public's opinions on certain vaccines \citep{Becker2016-uw, Blankenship2018-fl, Broniatowski2018-xv, Dunn2015-cu, Dunn2017-mf, Keim-Malpass2017-qj}. For example, \citet{Becker2016-uw} have analyzed the contents of tweets (primarily posted by users in India, Indonesia, and Vietnam) containing sentiments regarding the pediatric pentavalent vaccine (DTP-HepB-Hib) \citep{Ministry-of-Health-and-Family-Welfare-Government-of-India2012-qj} programs in those areas. They found that 37\% of the tweets contained negative sentiments, while 63\% were positive or neutral; they also indicated that most of the Tweets contained links to websites or additional resources and did not add any additional content or comments. \citet{Blankenship2018-fl} took this process a few steps further—tweets are not only analyzed for their sentiments about vaccines, but also for their amount of engagement (retweets), categorizations of their content, and the types of curators that posted them. Results found no discernible variation in the number of times anti-vaccine tweets were retweeted across content categories. Twitter (12.9\%), content curator ``Trap It" (3.4\%), and the Centers for Disease Control and Prevention (1.9\%) were the top 3 domains among links in pro-vaccine tweets. Additionally, social media sites, including Twitter (14.9\%), YouTube (8.4\%), and Facebook (3.4\%), were the most prevalent among the links in anti-vaccine tweets. The most frequently occurring theme in tweets with the hashtag \#vaccineswork was the childhood vaccination (40\%). Vaccines could reduce outbreaks and deaths, according to 29\% of tweets, which also referenced worldwide immunization efforts and improvement 21\% of the time \citep{Blankenship2018-fl}. Similarly, other studies \citep{Broniatowski2018-xv} have been conducted to determine which types of accounts are most problematic in spreading misinformation on Twitter. An analysis of sophisticated bots, Russian trolls, and content polluters found that, regarding tweets about vaccines, Russian trolls can ``amplify both sides" to create an online public discourse that can undermine public health; sophisticated bots, which are designed to look like legitimate accounts, can further undermine public health by increasing the number of those who hold apparent anti-vaccine sentiments.

Other studies \citep{Dunn2015-cu, Dunn2017-mf} explored the possibility of whether users are more likely to post anti HPV (Human Papillomavirus) vaccine tweets after being exposed to them themselves.  Dunn et al. found that the probability of tweeting something negative after being exposed to negative tweets was 37.78\%, which was substantially higher than the likelihood of doing so after previously being exposed to neutral or positive tweets, which was 10.92\% \citep{Dunn2015-cu}. In a subsequent study, \citet{Dunn2017-mf} expanded these results by attempting to create a model to explain the variance in HPV vaccine coverage. The study utilized an abundance of variables such as the exposure to HPV vaccine information on Twitter, socio-economic factors (e.g., poverty, education, insurance), racial and ethnic composition, and geographic location. They found that opinion exposure about the HPV vaccine on Twitter had more sway in determining vaccine coverage than socioeconomic factors. 

Even more shocking, though, is the sheer volume of tweets containing information about vaccines—it has also been found that most of the tweets are posted by ordinary accounts (or lay consumers - i.e., not an academic, institutional, or celebrity), and when sources are linked in tweets, it is also generally a link to a post made by an ordinary account \citep{Keim-Malpass2017-qj}. It is generally agreed upon by these researchers \citep{Keim-Malpass2017-qj} that Twitter can be an effective way to monitor opinions about public health issues and disseminate accurate information about the same issues. However, given the polarity and divisiveness of the current Twitter climate, and the sheer volume of tweets being sent out, Twitter itself and the overall information landscape must be improved (i.e., fact-checking, monitoring problematic accounts, improving overall information and media literacy standards, etc.) before that goal can become a reality.

Other researchers chose to explore this dynamic of diverse public opinion as it exists on a different platform - YouTube \citep{Basch2017-ac, Briones2012-oc, Ekram2019-an}. For instance, \citet{Basch2017-ac} viewed and categorized (by poster) 87 YouTube videos containing the phrases ``Vaccine Safety" and ``Vaccines and Children." The three most common categorizations of video posters were ordinary consumers, internet or TV news, and individual health professionals; shockingly, 65.5\% of the videos were deemed to be ``anti-vaccine". Similarly, after analyzing 172 YouTube videos related to the HPV vaccine for their tone and sentiment, response and reaction, and video source, \citet{Briones2012-oc} found that more than 51\% YouTube videos containing negative HPV vaccination sentiments compared to 32\% of positive ones, and the ``anti-vaccine" videos are far more likely to be liked or shared than the positive videos. Conversely, another study \citep{Ekram2019-an} of HPV vaccination sentiments on YouTube found that whether the video was positive or negative did not influence how many shares or views it received. The study also found that most videos could be classified as anti-vaccine. Other researchers \citep{Covolo2017-ji} have conducted similar analyses in other languages and geographic locations. For example, 123 Italian YouTube videos about vaccines were analyzed, and the researchers discovered that 50\% of the videos were positive in nature, 23\% were negative, and 27\% were neutral. Additionally, the study notes that both negative and positive videos alike utilized a ``fear appeal" at a higher rate than any other persuasive strategy like solidarity, economic interest, etc. YouTube videos posted regarding vaccines (both positive and negative) are rooted in fear and disdain for those with the opposite opinion. This further emphasizes the detrimental impacts on public perception due to the dire state of online information seeking and sharing trends.

\subsection{COVID-19 Misinformation Studies} \label{sec:covid19misinformationstudies}
COVID-19 is the first global pandemic in the social media era. This new experience opened up many nuances of social media and the fight against misinformation and fake news. Social media platforms have features such as automated bots that can facilitate the spread of misinformation \citep{Chu2010-wb}. Specifically, malicious activities have increased to an unprecedented level on social media during this pandemic \citep{Depoux2020-ay}. The volume of COVID-19 misinformation led to dire consequences for the public and caused frontline workers to face even more challenges in stymieing the spread of coronavirus. Public health agencies called this unchecked volume of mis/disinformation on social media platforms - infodemic \citep{Islam2020-wl, Organization2020-mq}.

\citet{Kim2020-cg} examined the effects of exposure to misinformation during the COVID-19 pandemic and identified that exposure to misinformation reduces the need to seek more preventative and treatment information, making it difficult to curb the spread of the disease.

Researchers have been scrambling to keep up with the dissemination of misinformation. \citet{Islam2020-wl} collected articles from various online sources like fact-checking websites, social media, newspapers, and television networks to examine rumors, stigma, and conspiracy theories about COVID-19 and how they potentially impact individuals and communities. \citet{Lazer2020-oc} examined tweets by 1.6 million registered voters in the United States to determine who is sharing the misinformation and its sources. They determined that there is a strong political divide for sharing misinformation, and mostly shared by people over 50. They also found that the belief in misinformation is more prevalent in the younger population. 

\citet{Evanega2020-ki} investigated the topics spreading misinformation during the early parts of the COVID-19 pandemic. They found that the majority of the misinformation was driven by ``miracle cure" topics and that prominent figures were the driving force in the spread of misinformation. They also noticed that only 16.4\% of the overall conversation is about fact-checking or correcting the misinformation.

After analyzing 43.3 million tweets, \citet{Ferrara2020-jy} found that automated social bots are used to disseminate misinformation and political conspiracy theories related to COVID-19. \citet{Al-Rakhami2020-ie} proposed a framework to use six different machine-learning algorithms to detect COVID-19 misinformation. They collected the data using Twitter API at the beginning of the pandemic and manually labeled the data to train the models.

Even the vaccination to prevent COVID-19 is being debated, and the misinformation is spread by the opponents of vaccination more frequently compared to the proponents \citep{Jamison2020-qs}. Although officials are taking steps to handle the misinformation regarding the vaccine \citep{Cornwall2020-gz}, the efforts are still falling short \citep{Wardle2021-jr} to tackle the diverse reasons \citep{Loomba2021-rk} for the spread of misinformation.

\citet{Thelwall2021-fg} in their study found that while the majority of the vaccine hesitancy in the English language twitter-sphere is related to right-wing conspiracy, there is a significant minority (18\%) who are refusing the vaccine for non-political reasons like fear of being targeted as black, development and approval speed, etc. Their study implies that vaccine hesitancy is not just confined to right-wing echo chambers but can reach a wider audience.

\citet{Ahammad2024-pp} found that misinformation can also spread using a positive tone and usually promotes alternate medicine, healthy living, and natural remedies. The positive sentiment-based fake news often increases hope and confidence in the public and can, in turn, reduce caution and make it difficult to contain the spread of the virus. The author also found that the prevalence of negative news, usually focused on crime and justice, can reduce public trust in authorities and increase anxiety, skepticism, and vaccine hesitancy.

\subsection{COVID-19 Vaccine Sentiment and Stance Detection} \label{sec:covid19vaccinesentimentandstancedetection}

Sentiment analysis is one of the major research areas in natural language processing (NLP) and can help us determine the overall perception of the population about any topic. Many researchers performed sentiment analysis on tweets. Some of the sentiment analysis research during the COVID-19 pandemic shines a light on how people are responding to the pandemic \citep{Venigalla2020-yt, Imran2022-hy}.

\citet{Dubey2020-ia} performed sentiment analysis on tweets from different countries between 11th and 31st March 2020. The researcher used the Syuzhet package \citep{Jockers2020-kw}, which classifies the tweets into eight different emotion categories. Within the data, Germany, France, the USA, and China showed balanced emotions between positive and negative tweets, while other countries showed a more positive attitude.

\citet{Manguri2020-yi} used the TextBlob python library, a Naive Bayes sentiment classifier model, on the tweets about COVID-19 for the week of 9th to 15th April 2020. The researchers found that people's reactions vary from day to day, and the majority of the tweets were neutral.

\citet{Liu2023-yw} coded the tweets about COVID-19 from six political leaders using a template analysis technique in five dimensions of populist political communication styles. Their study showed that during a crisis, populist communication styles can influence public adherence to government policies, and a combination of engaging and intimate populist communication styles performs best.

Stance detection is somewhat different than traditional sentiment analysis. While sentiment analysis can detect whether a text is positive, negative, or neutral, stance detection can classify someone's opinion as in favor or against a given target, which may or may not be present in the text, regardless of the emotion of the text \citep{Mohammad2016-ue}.

\citet{Augenstein2016-as} worked on detecting stance from tweets towards a target topic that is not present in the tweet. They showed that conditional Long-Short Term memory encoding is a suitable stance detection approach for an unseen target.

\citet{Dey2017-ht} proposed a two-phased Support Vector Model (SVM) approach for stance detection on Twitter data. In the first phase, they classified the tweets into ``neutral" and ``other" (non-neutral). Then in the second phase, they classified the non-neutral tweets into ``favor" vs. ``against." This method outperformed the state-of-the-art models.

\citet{Cotfas2021-uj} worked with tweets between November 9, 2020, and December 8, 2020, the month following the COVID-19 vaccine announcement, and found that the majority of tweets were in ``neutral" territory and tweets in ``favor" outpass ``against" stance towards the vaccine.

\citet{Poddar2021-xq} extended the work of \citet{Cotfas2021-uj} by analyzing tweets from pre-COVID and post-COVID on data ranging from January 2018 to March 2021. They identified the stance of users towards the COVID-19 vaccine and analyzed the topics they are tweeting to find a reason for the change in public stance.

\subsection{De/Motivation Studies and Vaccination Intent} \label{sec:demotivationstudies and vaccinationintent}

While the majority of research considers misinformation to be the primary culprit for vaccine hesitancy and worked to identify them \citep{Islam2020-wl, Kricorian2021-bf, Suarez-Lledo2021-uf}, there are many different factors like racial fear, stigma, economic constraints, distrust of government, and many more, that can discourage people from the vaccination \citep{Daly2021-nl, Kricorian2021-bf, Murphy2021-yp, Truong2022-qs, Guidry2021-nl}. Research suggests people are usually motivated by gain, altruism, or a protective attitude \citep{Finney-Rutten2021-al}. Protection motivation theory implies that the severity and susceptibility increase the vaccine intention \citep{Ansari-Moghaddam2021-qi}.

Human psychology research proposes that there can be intrinsic motivation, where people are motivated by internal realization, and extrinsic motivation, where external forces steer people towards something \citep{Oudeyer2007-hh}. And instead of competition, rewards, or threat of punishment, intrinsic motivation such as earning respect gains better results \citep{Deci1996-ai, Strickler2006-th}.

\citet{Schmitz2022-yu} support the previous results that autonomous or intrinsic motivation works best for vaccine intention and uptake, while controlled motivation (pressured by outside sources) does not work. They also noticed that people get more motivated by infection-related risk perception where personal health is at risk, rather than pandemic-related health concerns where the overall societal health is considered. They also found that distrust towards science also impacts the vaccine intention. 

To acquire intrinsic motivation towards vaccination, understanding of the vaccination and trust in the science are necessary. Lack of understanding of scientific findings, distrust towards politicians and involvement of the federal government, fast-tracking and emergency authorization, concern about financial profits and political motives, and misrepresentation of the severity of COVID-19 are the primary reasons causing the failure to motivate people for vaccination \citep{Finney-Rutten2021-al, Morris2021-sq}. The efforts to motivate people and increase their vaccine knowledge fall short for several other reasons including, but not limited to, unavailability of insurance reimbursement for consultation, lack of counseling, unavailability of vaccine during a clinic visit, and ease of getting an exemption \citep{Vasudevan2021-tz}.

Compared with previous studies, we analyzed Twitter data over a longer period of time, which covers both before and after the rollout of major vaccines in the US and around the world. We also extracted the motivating and demotivating topics resonating in the Twitter-sphere regarding the COVID-19 vaccine and also analyzed the spread of the topics based on geographic locations in the US. We then analyzed the public stance toward the COVID-19 vaccine and identified the topics driving those stances. We also grouped the topics based on the political landscape of each state to investigate whether the misinformation tactics differ based on the majority political view of the area.

Existing studies emphasize that social media is essential for disseminating healthcare information. However, the overwhelming prevalence of misinformation makes it difficult to educate the public, and exposure to misinformation alters public opinion and makes the work of healthcare professionals even harder. Our study identifies patterns in the misinformation and topics impacting public opinion. This study offers a path forward to overcome the challenge of disseminating appropriate information to the public.

Our work in this study is novel in that we analyzed people's stances over time to identify the de/motivational topics influencing their stances towards the COVID-19 vaccine. While different programs and campaigns \citep{Venigalla2021-fi, Hunt2022-cf} were launched to educate people about COVID-19 and encourage the general public to vaccinate, our work can help identify specific topics that are impacting public motivation and can help in future emergencies to reach the public cutting through the noise of misinformation and have the most impact.

\section{Data Collection} \label{sec:datacollection}

We have built a \textbf{Twitter Dataset} consisting of tweets and author information. Once we prepared the dataset, we used a machine-learning classifier to classify the tweets as motivating or demotivating, identify the stance of the tweets, and extract the most prominent topics in both motivating and demotivating classes. We also prepared several smaller ground truth datasets to train the machine-learning models. Table \ref{tab:datasets} lists the different datasets and their purposes.

\begin{table}[!h]
\centering
\small
  \caption{Datasets, sources, and their purpose}
  \label{tab:datasets}
  \begin{tabular}{ p{3cm}p{3cm}p{6cm} }
    \toprule
        Dataset & Source & Purpose\\
    \midrule
        Twitter dataset & \citet{Chen2020-am} & Primary dataset containing the tweets and author information. We classified this data and performed analyses on this.\\
    \hline
        Motivation Training dataset & \citet{Cheng2021-dc}, \citet{Muric2021-fn}, \citet{Brandwatch2020-wh} & Combination of three sources to build the ground truth dataset to classify de/motivating tweets.\\
    \hline
        Stance ground truth dataset & \citet{Poddar2021-xq}, \citet{Cotfas2021-uj}, manual labeling & Combination of three labeled datasets to build the ground truth dataset for COVID-19 vaccine stance detection.\\
    \bottomrule
  \end{tabular}
\end{table}

\subsection{Twitter Dataset} \label{sec:twitterdataset}
\begin{figure}[!h]
  \centering
  \includegraphics[width=0.7\linewidth]{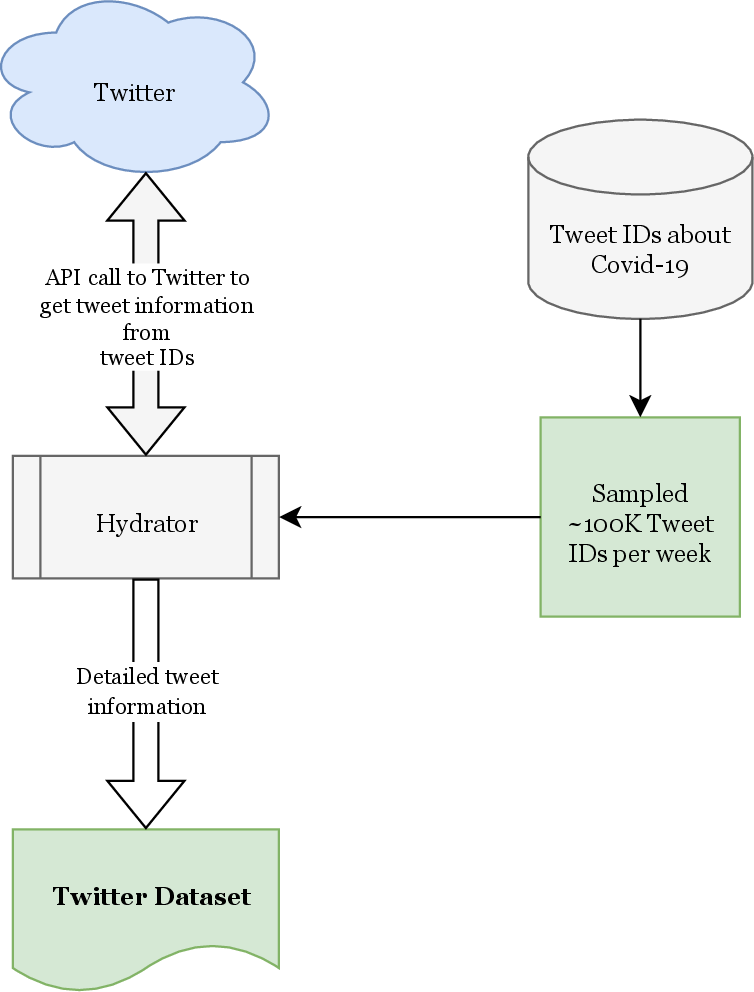}
  \caption{Preparing the datasets with COVID-19 related tweets.}
  \label{fig:dataset-prep}
\end{figure}

We have collected close to 16 million tweets between January 2020 and December 2021 that contain information about COVID-19. We have used the data from  \citet{Chen2020-am}  by gathering the Tweet IDs from their GitHub repository \citep{Chen2020-qm}. Chen et al. used several keywords like `Coronavirus,' `Corona,' `COVID-19,' `Pandemic,' `stayathome,' etc., to search for COVID-19 related tweets. In order to meet the rate limit of Twitter API \citep{Twitter2020-wu} and collect the data within a reasonable amount of time, we had to reduce the number of tweets. We randomly sampled at least 100,000 IDs each week and made sure that the data was stratified to match the distribution of the source dataset \citep{Chen2020-qm}. Then, we used the Hydrator API tool \citep{Documenting-the-Now2020-iy} to collect all the tweet information. We collected 15,768,845 tweets using this method. Figure \ref{fig:dataset-prep} shows the steps of the dataset creation.

\begin{figure}[!h]
  \centering
  \includegraphics[width=\linewidth]{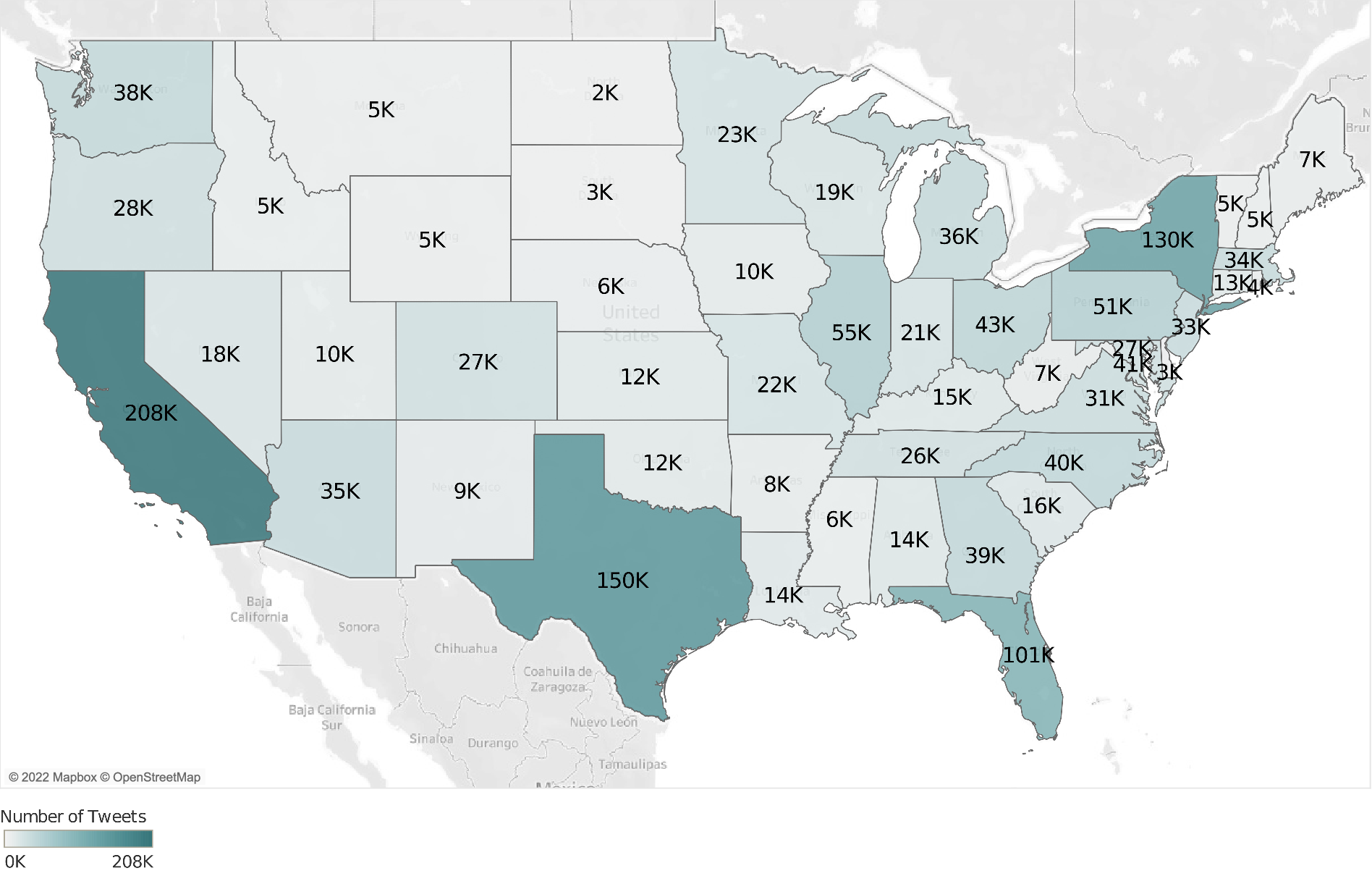}
  \caption{Number of tweets in different states (not showing Alaska, Hawaii, and distant territories).}
  \label{fig:usmap}
\end{figure}

After the data collection, we used the GeoPy \citep{GeoPy2020-in} library to get the geographic location of the users from OpenStreetMap API \citep{Nominatim2020-zt}. In this step, we only considered tweets in the English language. Based on the location gathered using the API, we isolated the tweets from the United States and labeled each tweet by respective US states and territories. We dropped the tweets without any geographic locations for the authors. For a few tweets, we manually corrected any mislabeling of states with the help of other available information in each tweet, such as zip code, landmark name, etc. At the end of the process, we had \(7,772,236\) tweets in our dataset. Figure \ref{fig:usmap} shows the distribution of tweets in major geographic locations in the US. Finally, we created a stratified set of \(466,335\) tweets spread throughout the two years for our experiments. We ensured the frequency of tweets per week represents the original \(7.7\) million data. This new set of tweets is the primary dataset for the study used for the classifications and analysis. 

We also performed a cleanup of the dataset by removing retweet tags (``RT"), newlines (``\textbackslash{n}"), special characters, URLs, and words that contain non-English characters. We analyzed the remaining tweets for duplicates and same-author duplicates. We noticed that there are very few duplicate tweets in the dataset. There are only \(0.37\%\) tweets that have more than ten duplicates, insignificant enough to cause any bias in the data. In the case of duplicate tweets from the same user, we found only \(0.13\%\) had more than three duplicate tweets, and no user had more than six duplicate tweets. We did not remove these duplicates since they are small enough to cause any bias, and the duplication of tweets may contain signals about society's emotions, which can be useful.

\subsection{Motivation Training Dataset} \label{sec:motivationtrainingdataset}
We created a dataset to train our machine-learning models to classify the tweets in the Twitter dataset as motivating or demotivating. We combined data from three different sources to build a robust training dataset for our models. We combined the COVID-19 rumor dataset by \citet{Cheng2021-dc}, the ``Avax Tweets" dataset - a COVID-19 vaccine hesitancy dataset from \citet{Muric2021-fn}, and our own collection of authentic tweets about COVID-19 vaccination.

\subsubsection{COVID-19 Rumor Dataset} \label{sec:covid19rumordataset}

It is a labeled dataset that contains COVID-19 rumors from both news sources and Twitter. The authors \citet{Cheng2021-dc} manually labeled 6,834 data points (\(4,129\) rumors from news and \(2,705\) rumors from tweets). We used the texts of the rumor and the label indicating whether the text is true, false, or unverified from this dataset. Previous studies \citep{Kim2020-cg, Xin2021-oo, Lee2022-ja} suggest that exposure to inaccurate news and misinformation reduces vaccine intent. But more accurate information and interpersonal communication motivate towards vaccination. Following the findings of these studies, we considered the ``true" news and tweets as motivating for vaccination while the ``unverified" and ``false" as demotivating.

\subsubsection{Avax Tweets Dataset} \label{sec:avaxtweetsdataset}

The authors \citet{Muric2021-fn} curated a list of tweets that exhibit an antivaccine stance. The dataset contains over \(1.8\) million tweets over one year, from October 2020 to November 2021. We have created a stratified sample of \(100,000\) tweets from the dataset, ensuring the frequency of tweets per week is representative of the original dataset. Then, using the Hydrator API tool, we collected \(79,093\) tweets from this sample. We considered these anti-vax tweets as demotivating tweets towards vaccination. We extracted the tweets from the dataset and labeled them as demotivating. 

\subsubsection{Authentic Vaccination Tweets Dataset} \label{sec:authenticvaccinationtweetsdataset}

We collected historical tweets regarding COVID-19 vaccines from a curated list \citep{Moore2020-pe} of trusted sources from Fortune magazine. The list contained trusted public health officials, epidemiologists, virus experts, family doctors, and prominent health organizations. The authors of these accounts shared their experience in treating patients during the COVID-19 outbreak, their advice, and refuting misinformation. This curated list gives us a source of authentic tweets regarding the pandemic and vaccination that are actively motivating people to vaccinate and advising the best ways to stay safe. We used the \citet{Brandwatch2020-wh} API to collect COVID-19 vaccine-related tweets by the users in the aforementioned list between January 2020 and December 2022. We collected \(19,992\) tweets using this process, extracted the tweet texts, and labeled the tweets as motivating.\\

We merged the three datasets above to create the ground truth dataset for training our models. We labeled the dataset with binary classes indicating whether a tweet contains antivaccination rhetoric (demotivating) or not (motivating). This will be our ``\textbf{motivation training dataset}" containing two features - the text and the label. Before training the machine-learning models, we cleaned the tweets by removing duplicates, retweet tags (``RT"), newlines (``\textbackslash{n}"), special characters, URLs, and words that contain non-English characters. Finally, we converted all the tweets to lowercase. After the cleanup, our dataset contained \(60,647\) demotivating tweets and \(21,235\) motivating tweets. We upsampled the motivating tweets to create a balanced dataset with \(121,294\) entries.

\subsection{Stance Ground Truth Dataset} \label{sec:stancegroundtruthdataset}

We extended the work of \citet{Poddar2021-xq} and \citet{Cotfas2021-uj} to identify the stance of tweets on the topic of vaccination. Although Poddar et al. \cite{Poddar2021-xq} published their trained model, it did not perform well with newer tweets. After manually checking their results, we found that the stance prediction was correct for \(55\) and \(61\) percent for anti and pro-vaccination, respectively. We believe this resulted from the model being trained with data from a smaller timespan. Therefore, we decided to ignore their model, use their labeled data in combination with our own, and train a machine-learning model with this newer data from a wider timespan. For this purpose, we manually labeled \(1,064\) tweets (\(469\) in favor, \(195\) against, and \(400\) unrelated) and combined that with the data from \citet{Cotfas2021-uj} (\(991\) in favor, \(791\) against, and \(1010\) unrelated) and the data from \citet{Poddar2021-xq} (\(1,364\) in favor, \(490\) against, and \(1,285\) unrelated). Finally, our ground truth data contained \(6,995\) entries with \(2,824\) in favor, \(1,476\) against, and \(2,695\) neutral tweets for the COVID-19 vaccine.

While manually labeling our data, we used two annotators to ensure there was no bias in the labeling. The Cohen's Kappa score for the two annotators was \(0.624\), meaning the labeling of the two annotators aligns at a satisfactory level. We also manually checked \(100\) random tweets from Cotfas et al. and \(200\) random tweets from Poddar et al., and our labeling aligned more than \(80\%\) of the time.

We have added a comprehensive list of all the data sources and their purpose in the \ref{sec:explanationofdatasetsused}.

\section{De/Motivating Topic Identification} \label{sec:demotivatingtopicidentification}
\begin{figure}[!h]
  \centering
  \includegraphics[width=0.8\linewidth]{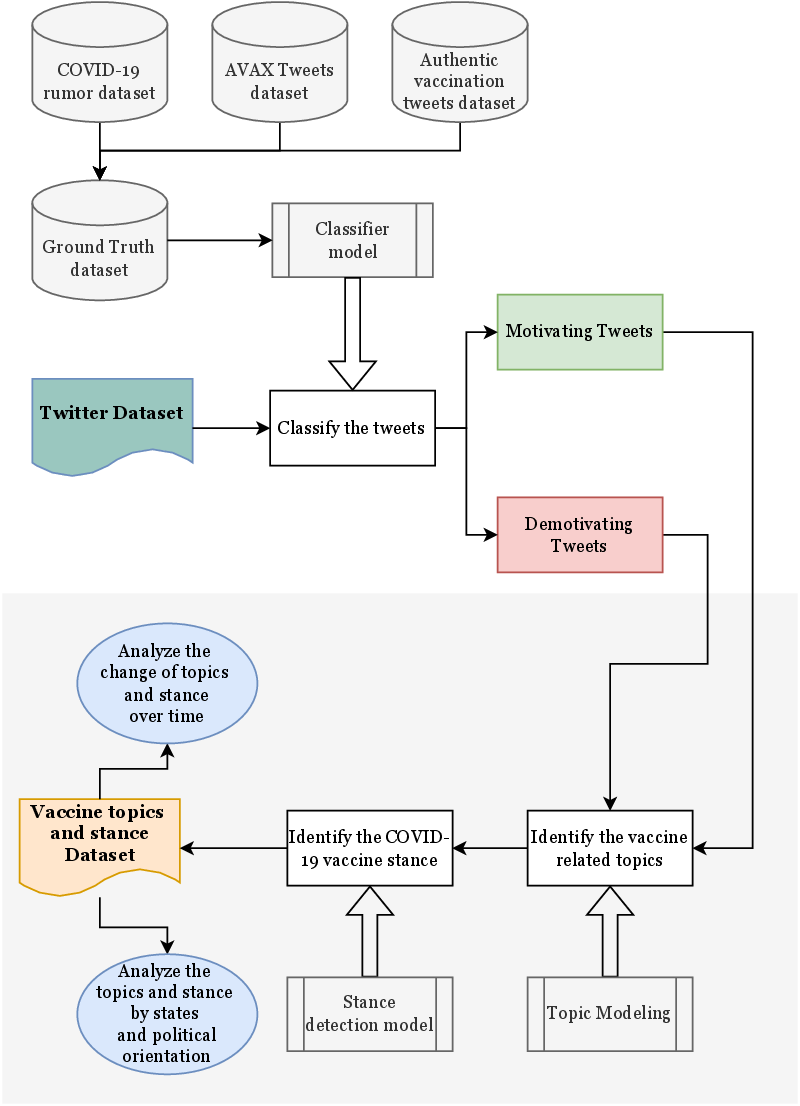}
  \caption{Classification of tweets and identifying the topics and stance.}
  \label{fig:tweet-classification}
\end{figure}

We have used machine-learning models to classify the tweets from our Twitter dataset as motivating or demotivating. Then, we used topic modeling to identify the prominent topics related to vaccines within the classified tweets. Figure \ref{fig:tweet-classification} shows the steps of our analysis.

\subsection{Classifying the Tweets} \label{sec:classifyingthetweets}

We fine-tuned DistilBERT \citep{Sanh2019-xt} and RoBERTa \citep{Liu2019-fh}, two different BERT-based pre-trained Natural Language Processing (NLP) models from the ``Hugingface transformers library" \citep{Wolf2020-bp} and found DistilBERT to be performing the best. We tried different learning rates, batch sizes, and epochs to measure performance. With a 70-30 train-test split, the accuracy for DistilBERT was \(96.3\%\). Table \ref{tab:classifiers} shows the performance of our training models.

\begin{table}[!h]
\centering
\small
  \caption{Models trained to identify de/motivating tweets}
  \label{tab:classifiers}
  \begin{tabular}{ p{0.5\columnwidth}c }
    \toprule
        Model & Accuracy\\
    \midrule
        DistilBERT & 96.3\%\\
    \hline
        RoBERTa & 73.4\%\\
    \bottomrule
  \end{tabular}
\end{table}

After the training, we used the model to classify the tweets in the Twitter dataset to identify the motivating and demotivating tweets. The DistilBERT model classified \(97,736\) as motivating and \(368,597\) as demotivating tweets. After we classified the tweets, we used topic modeling to identify the vaccine-related topics from each class.

\subsection{Topic Analysis} \label{sec:topicanalysis}

We used BERTopic \citep{Grootendorst2021-se} to create the topic models from our datasets. Although Latent Dirichlet Allocation (LDA) \citep{Blei2003-nx} is one of the most popular algorithms to do topic analysis, it takes some effort in hyperparameter tuning to generate meaningful topics and uses centroid-based topic extraction from document clusters. We also reviewed the Top2Vec \citep{Angelov2020-ag} topic modeling algorithm that trains the document and word vector jointly in a single semantic space. However, BERTopic leverages transformers \citep{Wolf2019-fg} and c-TF-IDF to create dense clusters allowing for easily interpretable topics. It is also possible to use pre-trained sentence transformer embedding models with BERTopic and find the prominence of any topic over time. Besides, the interactive visualization techniques make it much easier to investigate topic distribution using BERTopic.

We generated two separate topic models for the two classes in our dataset. After the training, each topic model returned the list of topics corresponding to the documents (tweets). We then extracted the top \(10\) topics related to ``vaccine" from each topic model.

\subsubsection{Motivating Topics} \label{sec:motivatingtopics}

We fitted the model exclusively with the tweets classified as ``motivating" to identify the motivating topics. The top 10 frequent topics from the model are displayed in Table \ref{tab:motivating-topics} with the number of tweets in that topic and the top 5 words in that topic. We have ignored the most frequent topic that contains stop words and pronouns.

Then, we extracted ``vaccine" related topics from the topic model. This process finds the topics similar to the keyword using cosine similarity. Table \ref{tab:motivating-vaccine-topics} shows the top 10 ``vaccine" related topics from the motivating tweets. The score in the table represents the topic's semantic similarity with the keyword ``vaccine".

\begin{figure}[!h]
  \centering
  \includegraphics[width=0.8\linewidth]{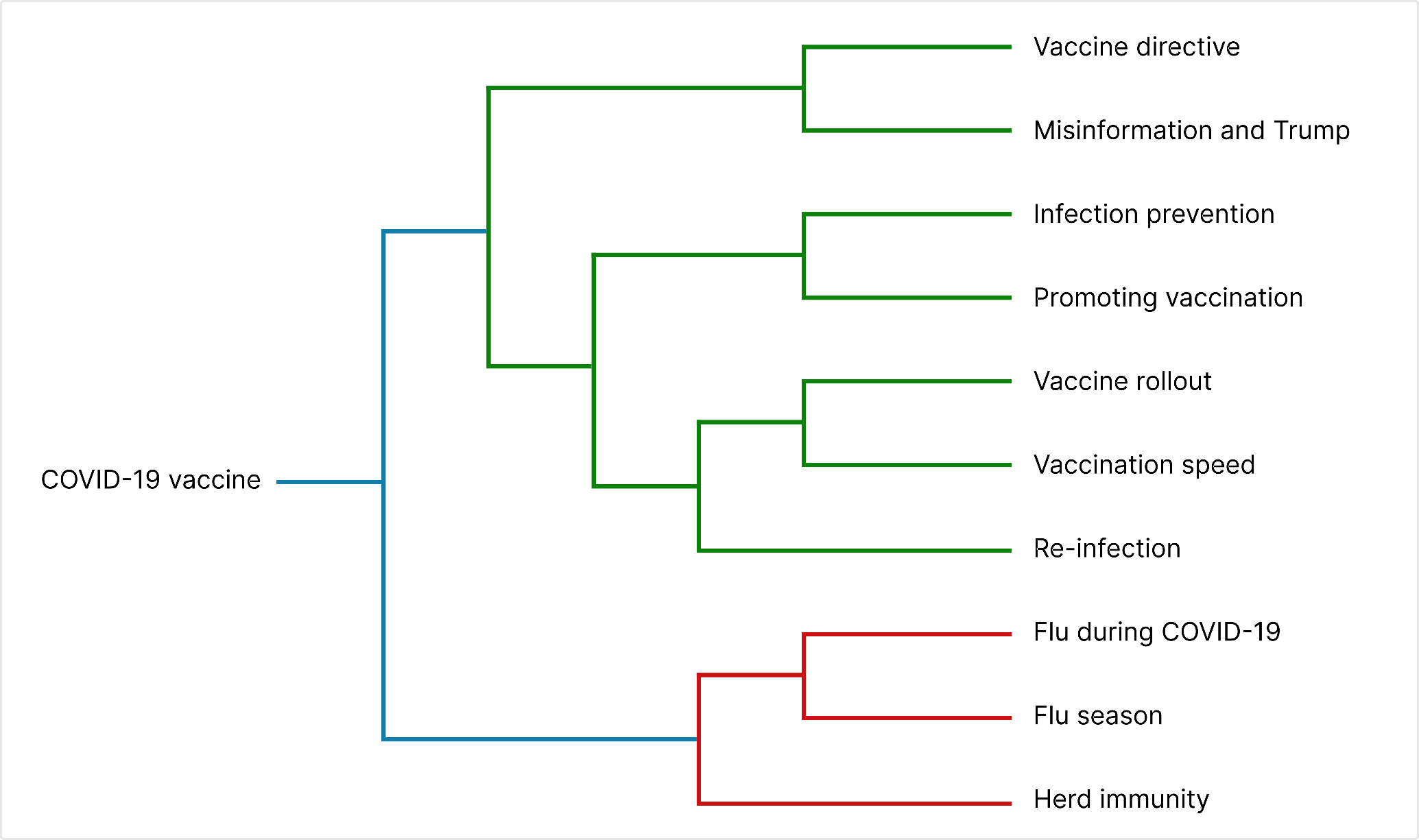}
  \caption{Hierarchy of topics related to vaccine in motivating tweets.}
  \label{fig:motivating-topic-hierarchy}
\end{figure}

\begin{figure}[!h]
  \centering
  \includegraphics[width=0.9\linewidth]{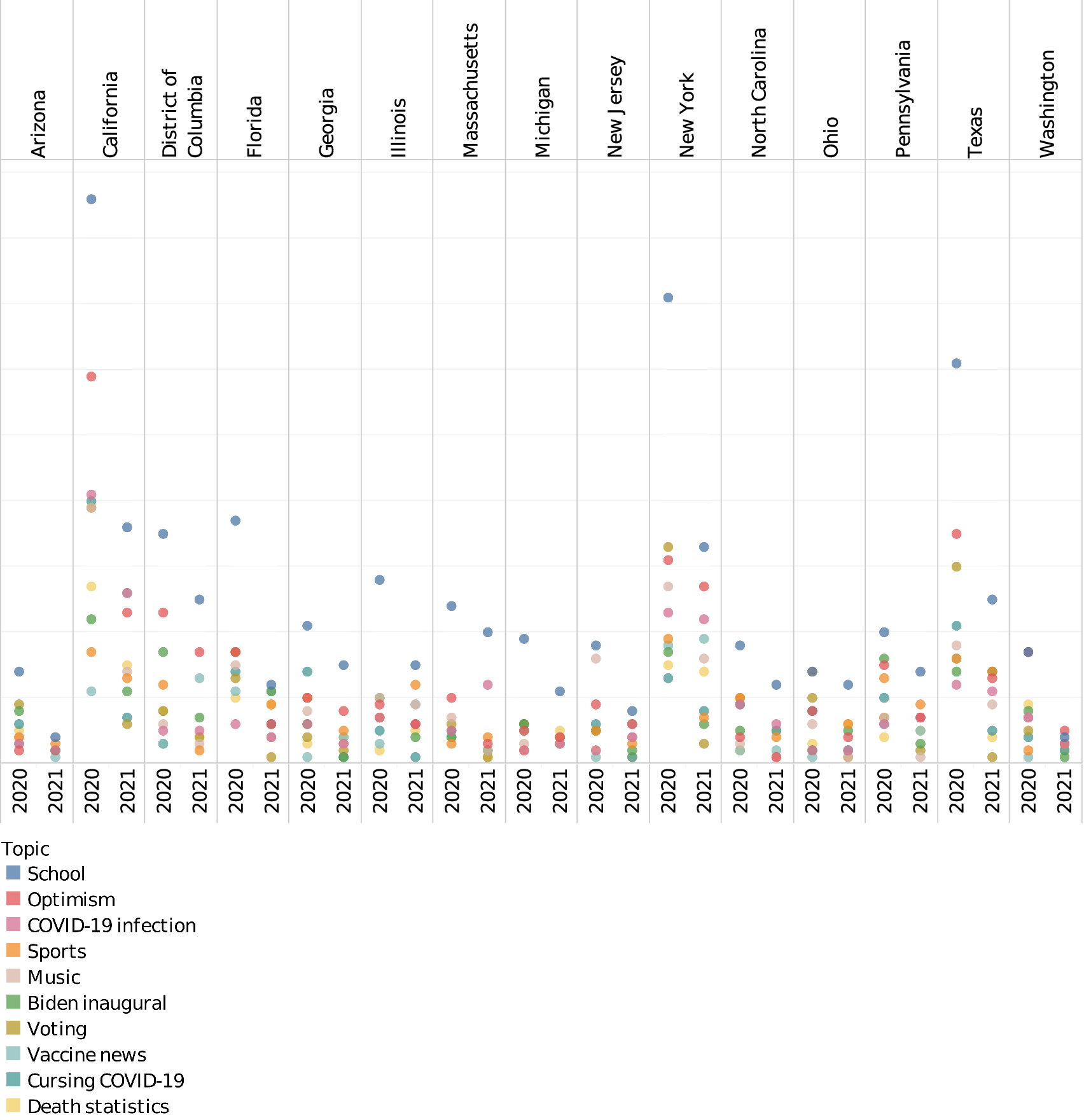}
  \caption{Top 10 topics in motivating tweets and their spread in the top 15 (by the number of tweets) US states.}
  \label{fig:motivating-topic-spread}
\end{figure}

The topic hierarchy in Figure \ref{fig:motivating-topic-hierarchy} shows the relationship between topics in the set, and Figure \ref{fig:motivating-topic-spread} shows the top 10 topics each year and the frequency of tweets for each topic.

\begin{figure}[!h]
  \centering
  \includegraphics[width=0.8\linewidth]{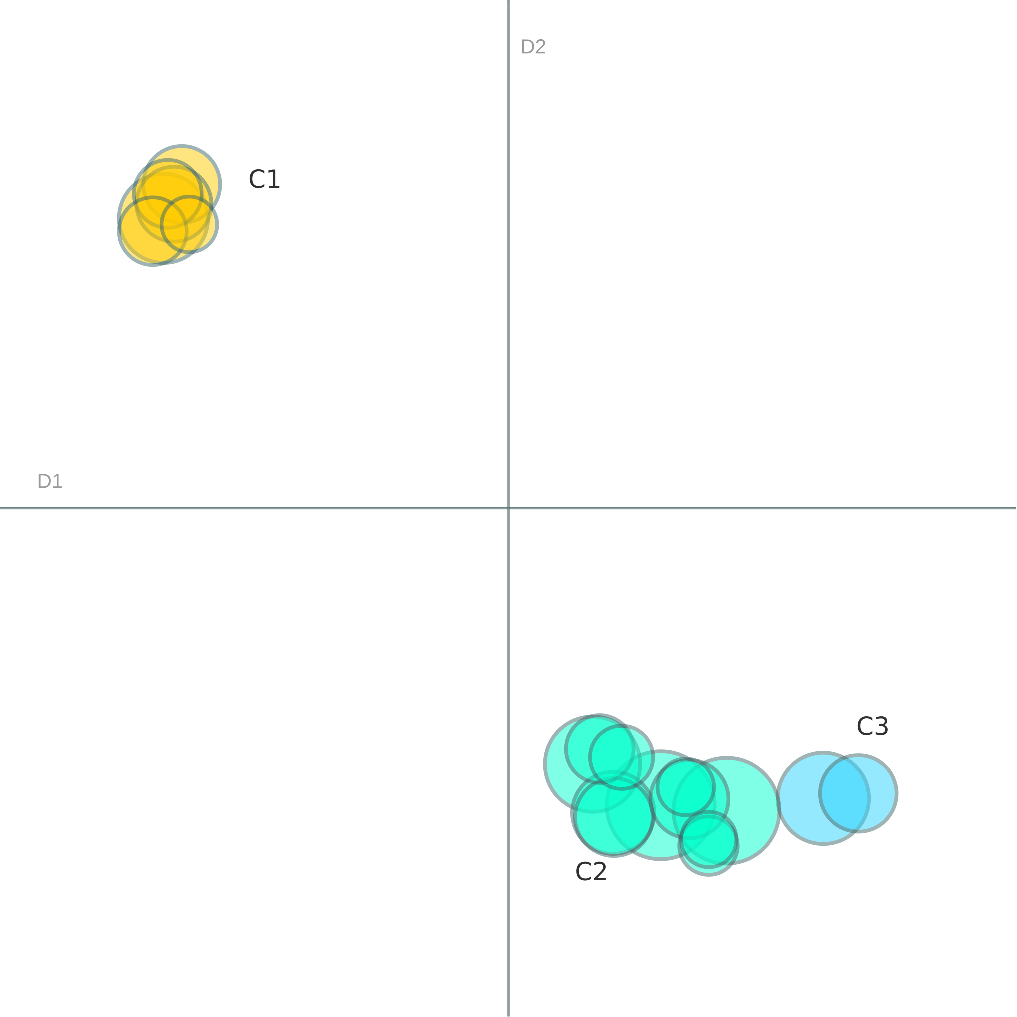}
  \caption{Cluster of motivating topics related to vaccine. In this figure, cluster C1 includes topics related to politics, cluster C2 includes topics related to vaccine development, and cluster C3 includes topics related to the COVID-19 task force.  Each bubble represent one topic and the size represents the number of tweets in that topic.}
  \label{fig:motivating-topic-cluster}
\end{figure}

Clustering the motivating topics related to ``vaccine" shows that the topics are mostly clustered in two regions as displayed in Figure \ref{fig:motivating-topic-cluster}.

\subsubsection{Demotivating Topics} \label{sec:demotivatingtopics}

We performed the same steps with demotivating tweets. We fitted the topic model with tweets classified as ``demotivating." Table \ref{tab:demotivating-topics} shows the top 10 frequent topics from the model, the number of tweets in each topic, and the top 5 words in that topic. Pronouns and stop words are ignored as before.

Similar to the earlier model, we extracted ``vaccine" related topics from the topic model. Table \ref{tab:demotivating-vaccine-topics} shows the top 10 ``vaccine" related topics from the demotivating tweets.

\begin{figure}[!h]
  \centering
  \includegraphics[width=0.8\linewidth]{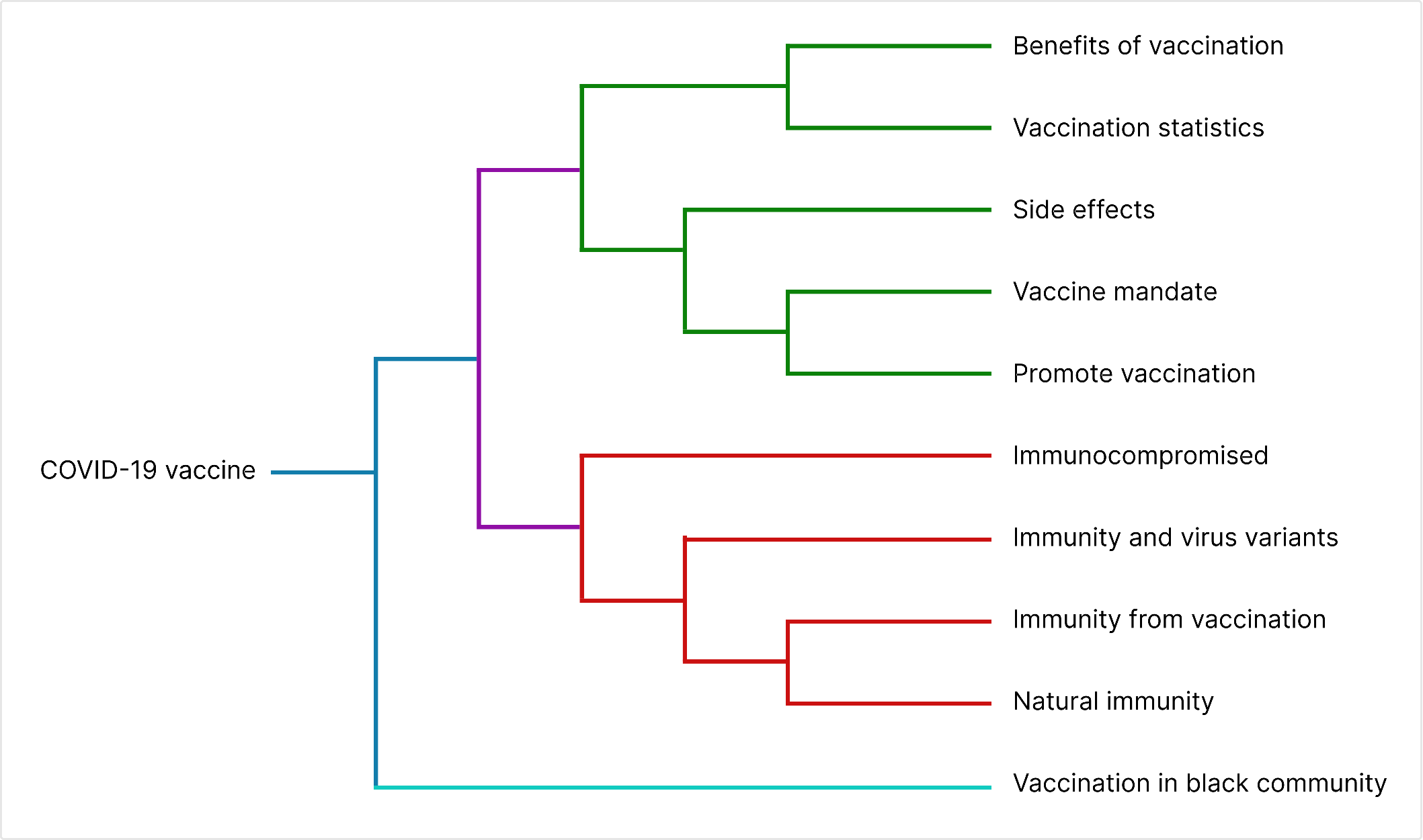}
  \caption{Hierarchy of topics related to vaccine in demotivating tweets.}
  \label{fig:demotivating-topic-hierarchy}
\end{figure}

\begin{figure}[!h]
  \centering
  \includegraphics[width=0.9\linewidth]{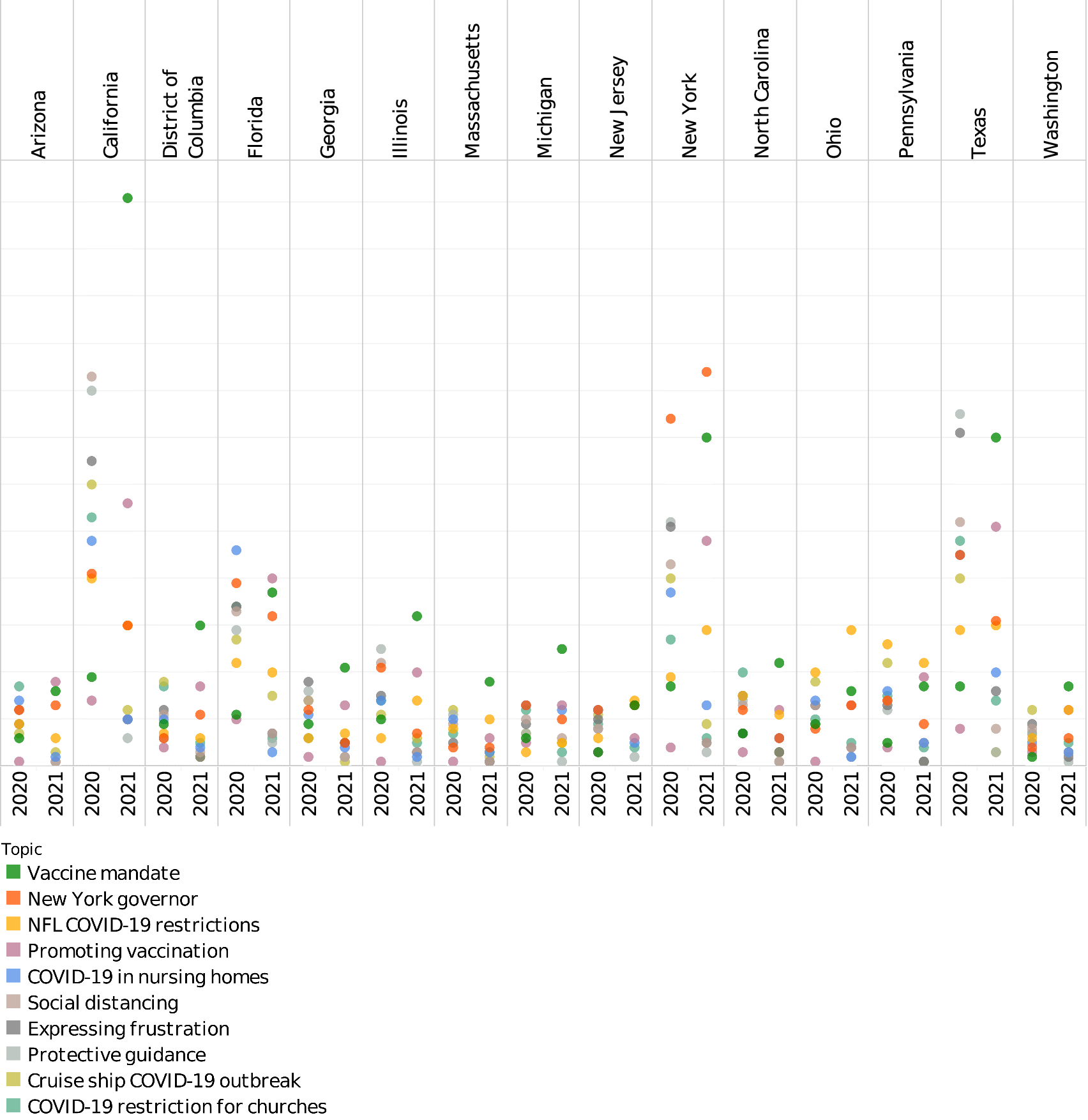}
  \caption{Top 10 topics in demotivating tweets and their spread in the top 15 (by the number of tweets) US states.}
  \label{fig:demotivating-topic-spread}
\end{figure}

The topic hierarchy in Figure \ref{fig:demotivating-topic-hierarchy} shows the relationship between topics in the set. Figure \ref{fig:demotivating-topic-spread} shows the top 10 topics each year and the frequency of tweets for each topic.

Clustering the demotivating topics related to ``vaccine" shows that the topics are scattered in different clusters, as displayed in Figure \ref{fig:motivating-topic-cluster}, which is different from the motivating topics.

\begin{figure}[!h]
  \centering
  \includegraphics[width=0.8\linewidth]{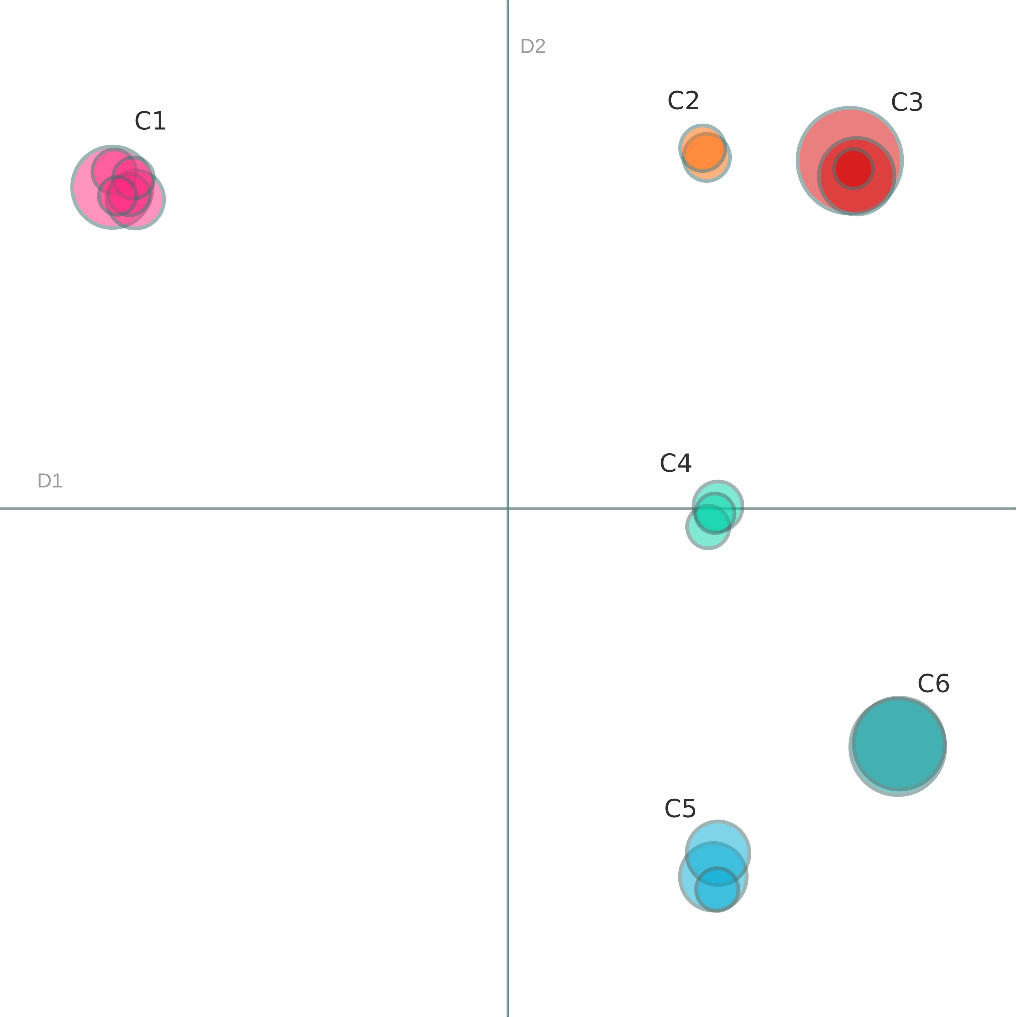}
  \caption{Cluster of demotivating topics related to vaccine. In this figure, clusters C1, C2, C3, C4, C5, and C6 includes topics related to healthcare and medication, liberal politics, policies, government agencies, misinformation, and conservative politics respectively. Each bubble represent one topic and the size represents the number of tweets in that topic.}
  \label{fig:demotivating-topic-cluster}
\end{figure}

\section{Stance Detection} \label{sec:stancedetection}

The stance of a text tells us whether the author is in favor or against a topic. To find the stance of users for the COVID-19 vaccine, we trained machine-learning models using this labeled data discussed in section \ref{sec:stancegroundtruthdataset} for stance detection. We used SimpleTransformer \citep{Rajapakse2021-pr} to train models using Huggingface \citep{Wolf2019-fg} transformers. Table \ref{tab:stance-transformers} shows the transformer training results reporting the performance in Matthews Correlation Coefficient (MCC). We used the MCC to measure performance because this provides a better understanding of the performance on an imbalanced dataset \citep{Chicco2020-oq}. Since the ratio of pro-vaccine, anti-vaccine, and neutral are not balanced, MCC provides a better understanding than traditional accuracy or F1 score. We trained Roberta and COVID-Twitter-BERT (ct-BERT) \citep{Muller2020-bt} with different epochs. The best-performing model is highlighted in bold in Table \ref{tab:stance-transformers}, which is ct-BERT with 10 epochs and is referred to as ct-BERT-10 hereafter.

\begin{table}[!htb]
\centering
\small
  \caption{Transformer training performance}
  \label{tab:stance-transformers}
  \begin{tabular}{ p{0.5\columnwidth}c }
    \toprule
    Transformer & MCC\\
    \midrule
        Roberta (epochs: 10) & 0.490\\
    \hline
        CT-BERT (epochs: 5) & 0.580\\
    \hline
        \textbf{CT-BERT (epochs: 10)} & \textbf{0.603}\\
  \bottomrule
\end{tabular}
\end{table}

\begin{figure}[!h]
  \centering
  \includegraphics[width=\linewidth]{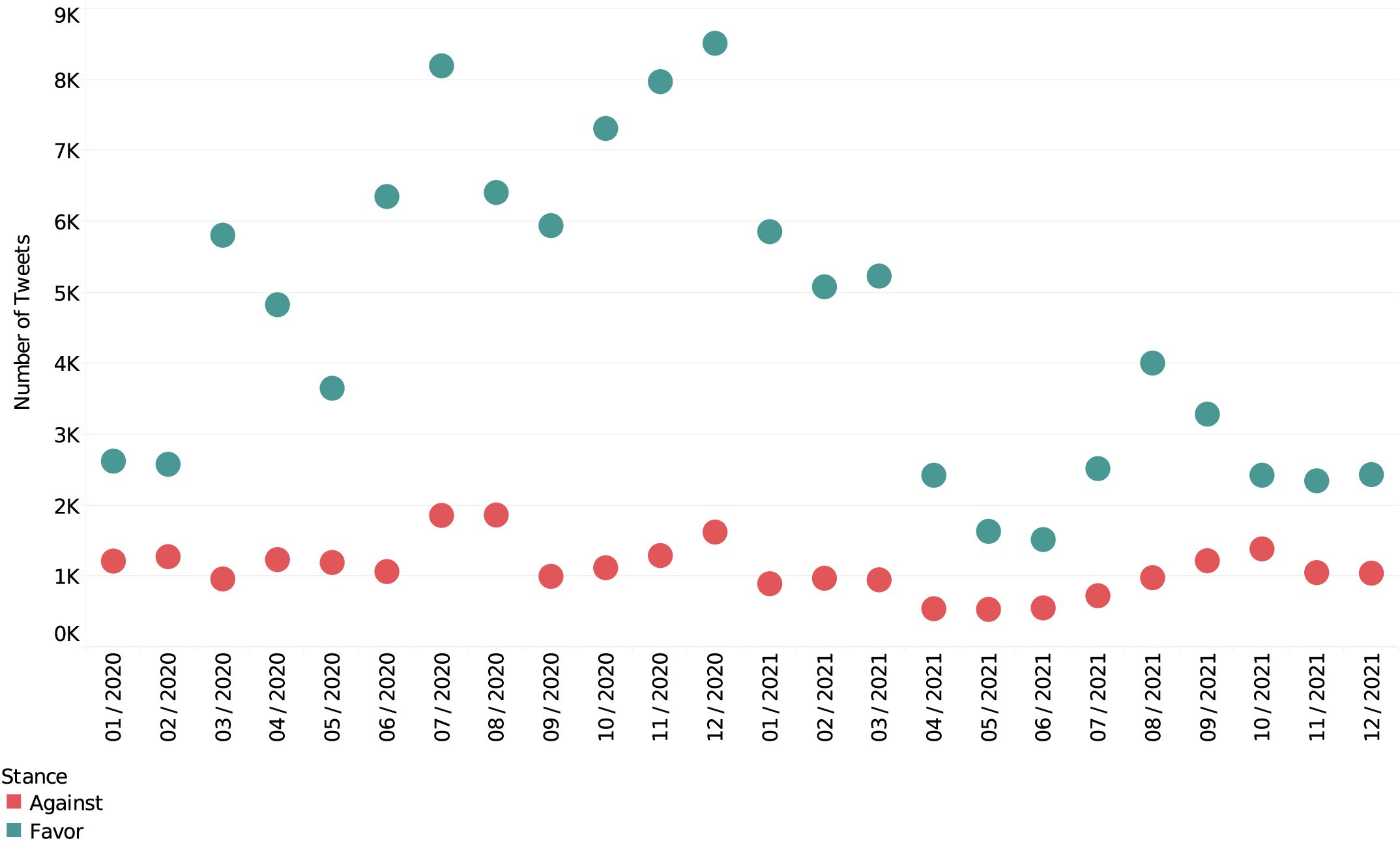}
  \caption{Stance towards vaccination (ignoring unrelated stance).}
  \label{fig:stance-per-month}
\end{figure}

\begin{figure}[!h]
  \centering
  \includegraphics[width=\linewidth]{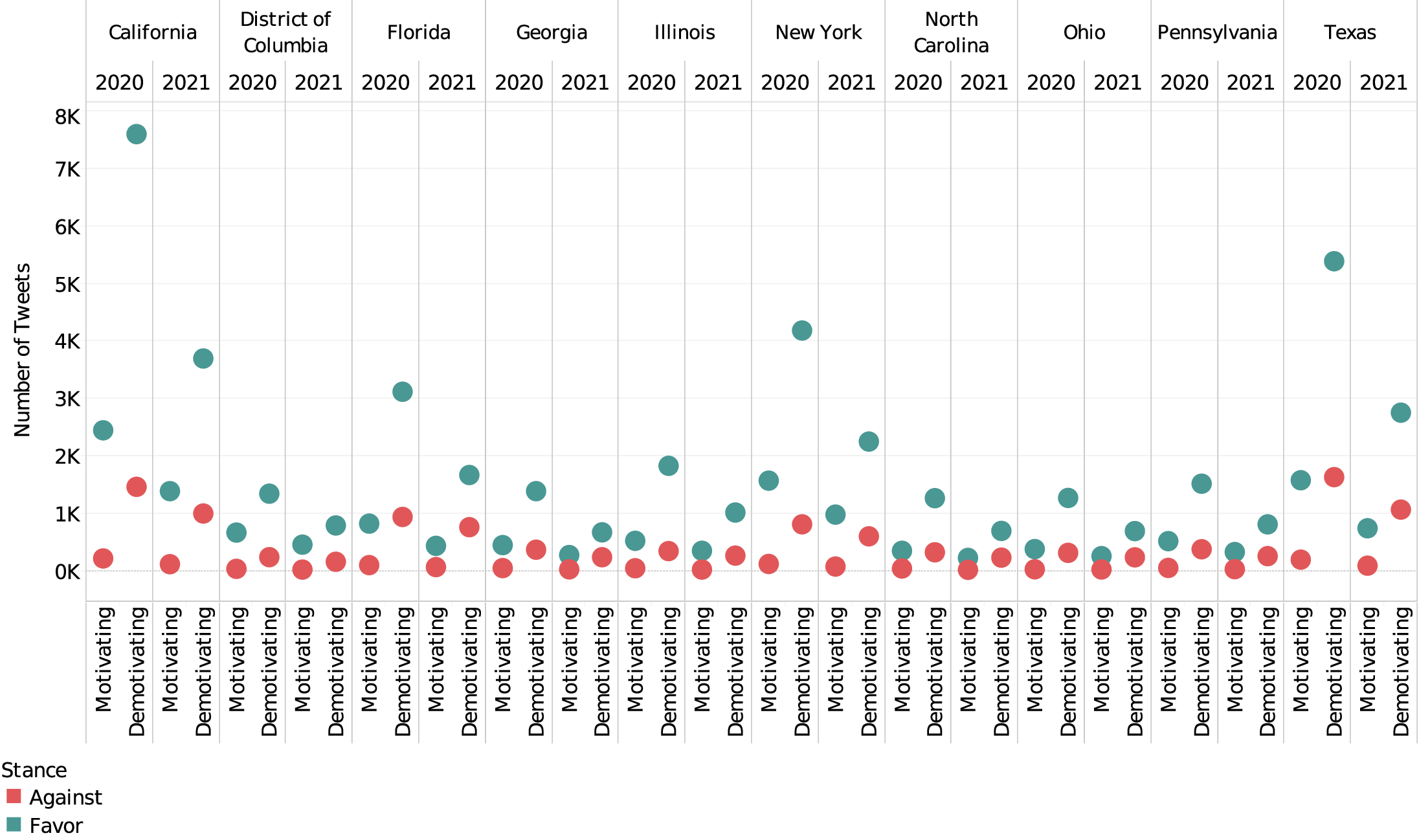}
  \caption{Vaccination stance in top 10 states by the number of tweets (ignoring unrelated stance).}
  \label{fig:stance-per-state}
\end{figure}

Then, we classified the Twitter dataset using the ct-BERT-10 model to identify the stance of the tweets. After classifying the tweets, we found that the `Favor' stance outpasses `Against' in different time segments and states. Figure \ref{fig:stance-per-month} shows the number of tweets in each stance each month, and Figure \ref{fig:stance-per-state} shows the stance in different states grouped by year and motivation classification. From Figure \ref{fig:stance-per-month}, we can notice some interesting trends, like spikes in the ``favor" stance in the middle and the end of the year 2020.

\section{Discussion} \label{sec:discussion}

The COVID-19 pandemic was a disrupting event, changing our lives in more ways than we can imagine. After over three years, we are still figuring out the impacts and learning to live with the new normal. This pandemic opened up new research horizons and questions about how we express ourselves in social media and how to handle a big challenge like this, motivate a large population towards specific activities, and prevent future disasters. 

Our research uncovered some fascinating trends. For example, the public stance on vaccination shifted depending on time and geographic location. We also found that using external motivators, like government mandates, could sometimes backfire and actually discourage people. This section goes deeper into these findings, addressing the research questions we outlined earlier. Table \ref{tab:sample-tweets} in \ref{sec:sampletweetsfromdifferentclasses} offers a sample of tweets from different motivating and demotivating categories and vaccination stances, providing a window into the broader dataset we analyzed.

\begin{figure}[!h]
  \centering
  \includegraphics[width=\linewidth]{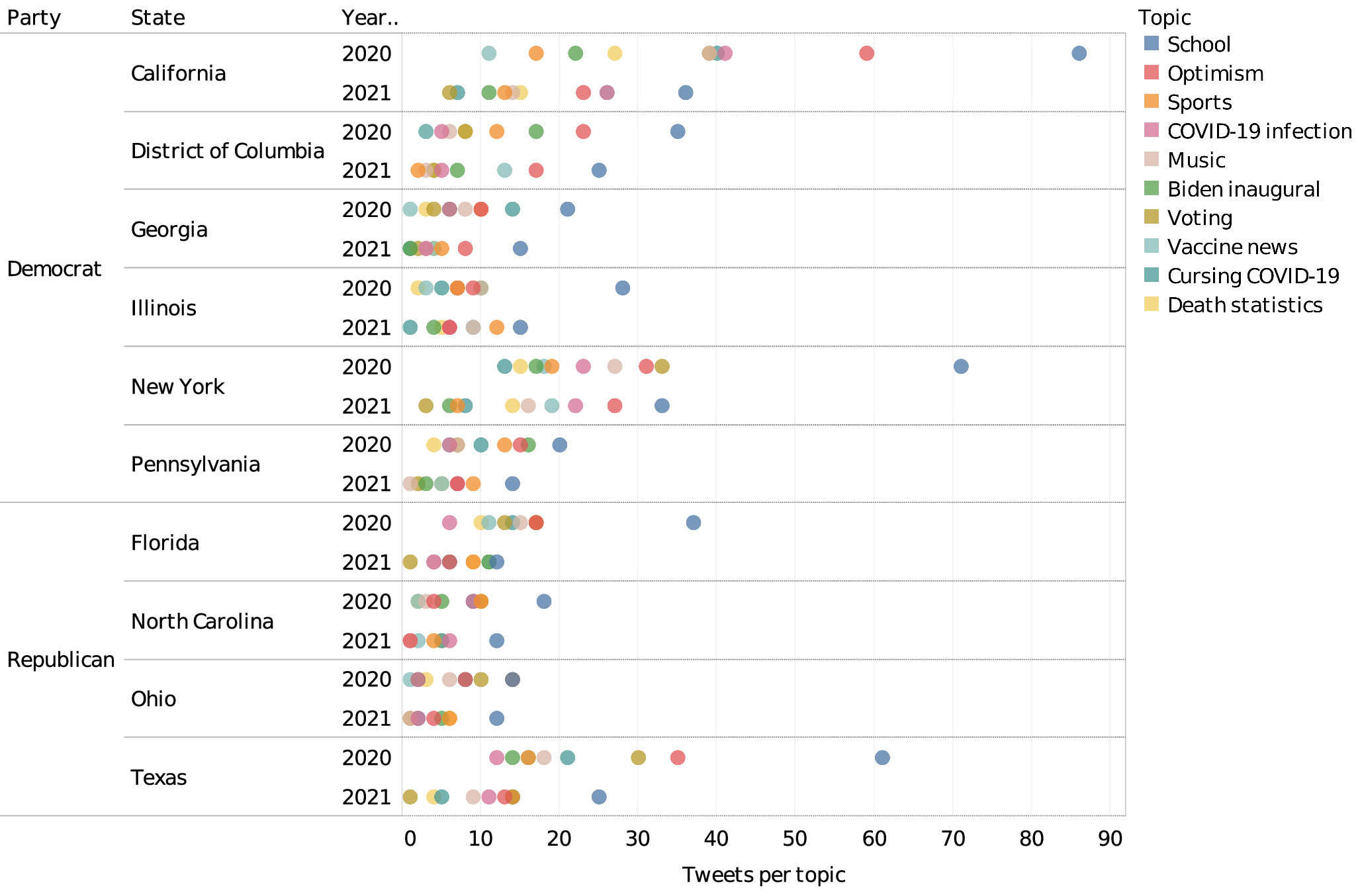}
  \caption{Motivating tweets in the top 10 states grouped by political affiliation in the 2020 presidential election.}
  \label{fig:motivating-topic-politics}
\end{figure}

Addressing our first research question (RQ\ref{rq:rq1}) necessitates establishing a framework for motivation and demotivation in the context of our study. Motivation is a subjective concept and a motivating statement for one individual might have the opposite effect on another. This study considers statements encouraging others to vaccinate for COVID-19 as motivating and vice versa.

Our analysis in section \ref{sec:topicanalysis} revealed that motivating tweets frequently address topics like schooling, voting, sports, music, and even COVID-19 statistics. Within these topics, concerns related to personal health, protective measures, and vaccine-related news emerged as motivational factors for vaccination.

Conversely, demotivational tweets often centered on themes of conservative political ideology, vaccine mandates, and protective directives from the officials. A deeper analysis of vaccine-specific topics within the demotivating category identified hesitancy regarding vaccine efficacy, concerns about potential side effects and immunity, and a sense of distrust towards the vaccine within minority communities.

\begin{figure}[!h]
  \centering
  \includegraphics[width=\linewidth]{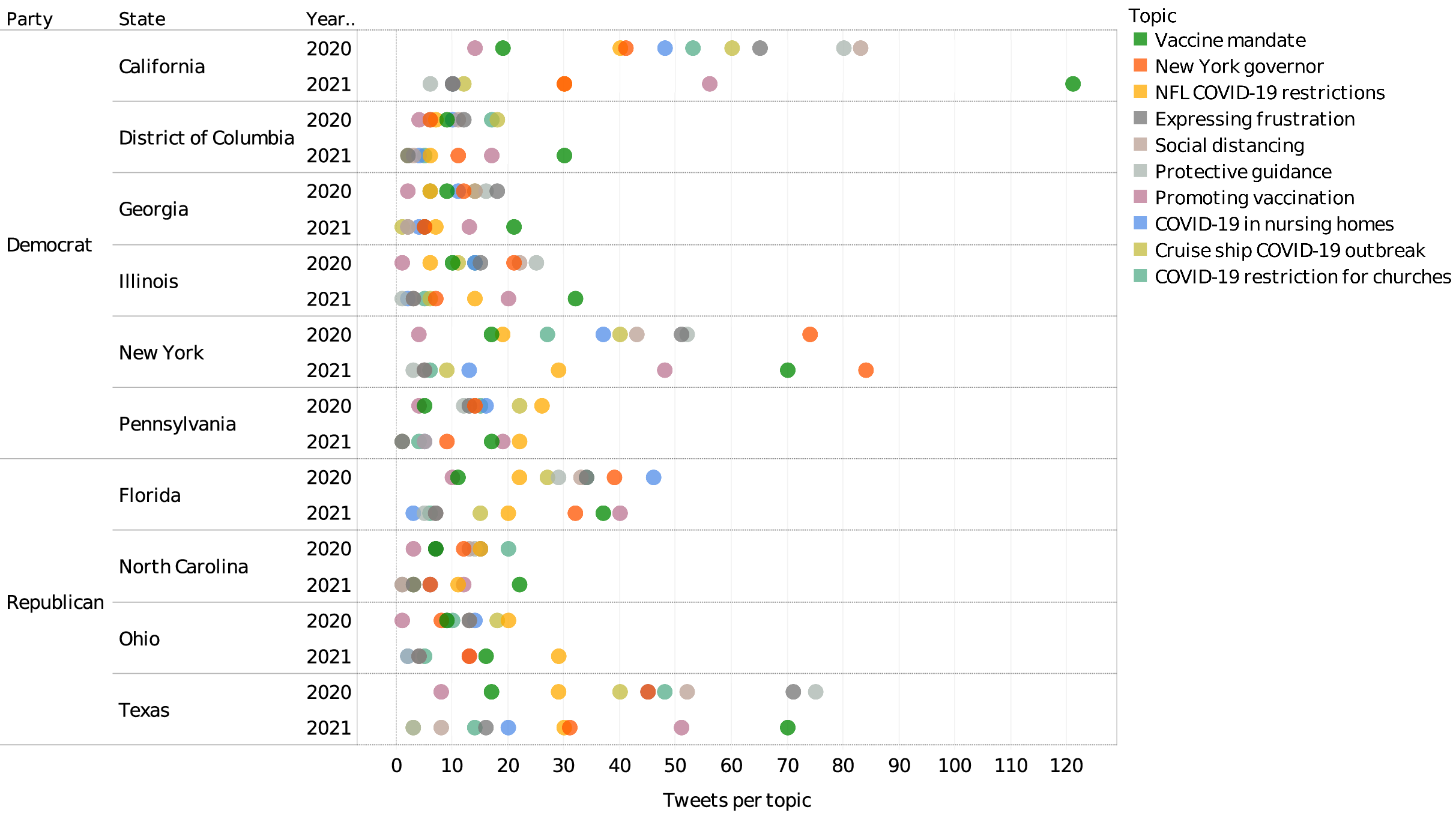}
  \caption{Demotivating tweets in the top 10 states grouped by political affiliation in the 2020 presidential election.}
  \label{fig:demotivating-topic-politics}
\end{figure}

A critical finding of this study is the considerable variation observed in demotivating topics across factors like political orientation and geographic location. In contrast, motivating topics remained largely constant, irrespective of these variables. This trend is evident from visualizations in Figures \ref{fig:motivating-topic-spread}, \ref{fig:motivating-topic-cluster}, \ref{fig:demotivating-topic-spread}, \ref{fig:demotivating-topic-cluster}, \ref{fig:motivating-topic-politics}, and \ref{fig:demotivating-topic-politics}.

Our study found minimal overlap between motivating and demotivating topics, suggesting the potential to identify and address specific concerns in hesitant communities. A key observation was that demotivating tweets demonstrated stronger political leaning and often towards conservative politics in the US. Interestingly, our analysis revealed that a subset of tweets promoting vaccination also demotivated the public.

Figure \ref{fig:motivating-topic-hierarchy} visualizes two prominent branches of discussion in the motivating tweets. One segment focuses on policy matters, vaccine development, and similar topics, while the other emphasizes the symptoms and the dire physical impacts of the disease to encourage vaccination. This observation aligns with the existing literature about intrinsic motivation and protective instincts associated with vaccine intent \citep{Ansari-Moghaddam2021-qi, Deci1996-ai, Schmitz2022-yu, Strickler2006-th}.

Similarly, Figure \ref{fig:demotivating-topic-hierarchy} illustrates three significant branches of discussion in demotivating tweets - vaccination policy and mandates in conjecture with vaccine promotion efforts, concerns surrounding the impact on immunocompromised individuals, and specific concerns and distrust in minority communities. Human psychology research suggests that extrinsic motivations do not achieve better results \citep{Deci1996-ai, Strickler2006-th}, which aligns with our observations. Additionally, the identified concerns in minority communities and the effects of vaccines on immunocompromised individuals highlight that there were factors beyond misinformation that contributed to vaccine hesitancy. This observation demonstrated the need for better and more nuanced messaging strategies from policymakers and healthcare workers to address these specific doubts and knowledge gaps within certain population segments.

With the branches identified from the study, our study offers the potential to tailor communication efforts to address specific concerns in different communities, encourage them to vaccinate, and dispel misconceptions. This process of identifying and addressing community-specific concerns can be invaluable for policymakers in future emergencies.

We explored our second research question (RQ\ref{rq:rq2}) about topics influencing public stance towards vaccination based on the stance detection analysis detailed in section \ref{sec:stancedetection}. Our findings suggest that a higher level of exposure to demotivating topics increases negative public stance. The visualization in Figure \ref{fig:stance-per-month} illustrates that the number of anti-vaccination tweets remained relatively stable over time, while tweets favoring vaccination fluctuated significantly. This observed fluctuation warrants further investigation to determine potential correlations with factors such as vaccine innovation news, emergence of new COVID-19 variants, or shifts in the political climate.

Figure \ref{fig:stance-per-state} also presents an interesting observation. Typically, for the majority of the tweets, demotivating tweets align with an anti-vaccination stance and vice versa, aligning with general expectations. However, there are also instances where opposing stances towards vaccination motivated and favorable stances towards vaccination demotivated the public. While seemingly counterintuitive, this finding aligns with research in human psychology \citep{Deci1996-ai, Strickler2006-th}, which suggests that excessive external pressure toward a specific action can lead to a phenomenon known as psychological reactance \citep{Steindl2015-bh}, where individuals resist the pressure and move in the opposite direction. Further study in this direction can give us more insight into the specific topics that are shifting the stance of individual users.

The analysis to answer our third research question (RQ\ref{rq:rq3}) about the trends in de/motivating topics over time and geographic locations found some compelling insights. As evident from Figure \ref{fig:motivating-topic-spread}, the top motivating topics related to COVID-19 vaccination remained relatively consistent across time and geographic locations. Conversely, demotivating topics exhibited significant variations over time and geographic locations, as shown in Figure \ref{fig:demotivating-topic-spread}. This disparity between motivating and demotivating topics is further evident in Figures \ref{fig:motivating-topic-politics} and \ref{fig:demotivating-topic-politics}, where motivating topics remained primarily consistent in different states regardless of their political orientation, while the demotivating topics exhibited variations.

We can extrapolate from our findings that regional politics played a crucial role in vaccine hesitancy during the COVID-19 pandemic and warrants finding a path to disseminate the message from policymakers and healthcare workers without politicizing them in future emergencies. This observation opens up new avenues for research into the impact of critical issues in regional politics during national or global emergencies.

Furthermore, the development of a visual analytics tool based on the results of this study \citep{Rahman2023-yh} offers an effective way to explore and communicate the results visually and help policymakers understand the trends in social media. There are numerous ways to understand the motivation of the masses and constructively communicate with them to disseminate a message. Our study underscores the importance of recognizing and comprehending the diverse motivations of different communities. Our work indicates that effective communication strategies, moving beyond generic messaging or simply refuting misinformation with brute force, are necessary to communicate with the public. Policymakers and healthcare workers should prioritize recognizing and addressing the unique concerns of specific populations to cultivate intrinsic motivation.

\section{Limitations} \label{sec:limitations}
We worked particularly with tweets related to COVID-19, spanning over two years. We acknowledge that nuanced factors were involved in public reaction during the pandemic. The social distancing requirements, stay-home orders, remote work, loss of jobs, and economic and political factors played important roles. There can be different sociopolitical scenarios in a future event that will require considering those while generalizing the work of this study.

We also acknowledge that the automated labeling of \textit{Motivation Training Dataset} in Section \ref{sec:motivationtrainingdataset} required the assumption of ``false" news and tweets as demotivating. While the assumption grounds itself on previous academic studies, future works with extensive user studies to validate the assumptions can strengthen the findings of the work and reveal new research areas.

In this study, we used a smaller sample of 500K tweets from the original 7.7M tweets to reduce resource usage. Implementing the same methods over the 7.7M tweets can yield further exciting observations. 

Furthermore, we did not dig into user-specific topics that impacted individual users to change their stance toward vaccination - which can be an interesting direction for future research.

\section{Ethical Concerns} \label{sec:ethicalconcerns}
There are some ethical concerns that we have taken into account. We anonymized the identifiable information on the tweets used for the machine-learning models and in the published datasets. Since we performed the anonymization programmatically, we acknowledge there is the possibility of a few tweets remaining identifiable based on the location or landmark information posted by the users in their tweets. We also acknowledge that malicious actors can also use the proposed method of reaching social media users and motivating them. Nevertheless, that is a predicament with all modern inventions and depends on who uses it. We believe the benefits of the findings in this study outweigh the risks.

\section{Conclusion} \label{sec:conclusion}
While existing research has primarily focused on identifying misinformation in COVID-19 tweets and performing sentiment and topic analysis, a critical gap exists in comprehensively understanding the specific topics that demotivated public opinion toward COVID-19 vaccination during the pandemic. This study addresses this gap by analyzing a large dataset of tweets spanning the early stages of the pandemic through the end of 2021. It was evident during the pandemic that the public trust in institutions and scientific processes had diminished. The study about the relationship of resonating topics on social media with public motivation can provide essential knowledge and tools to cut through the noise of misinformation and communicate with the public. This knowledge is essential to rebuild trust in the institutions and improve public understanding of scientific processes.

Our findings reveal a compelling link between social media topics and public vaccination motivation. This study demonstrates that factors beyond misinformation influence vaccine hesitancy. While misinformation remains a concern, this research highlights the importance of intrinsic motivation and protective instinct in driving vaccine uptake. Our analysis also underscores the critical role of local politics in shaping public opinion.

In the current era of social media, information is at the fingertip of the public. Furthermore, the wide availability of LLMs facilitates the threat of misinformation and disinformation masquerading as credible information circulated through the population. Without appropriate knowledge to critically evaluate information, the interpretation, often by non-specialists, makes it more dangerous and causes adverse outcomes. The politicization of messages diluted with misinformation and misunderstanding also makes it impossible to educate the people during emergencies.

Our work underscores the importance of comprehending human psychology to motivate people during emergencies. External incentives and excessive regulations from institutions often backfire. By contrast, communication strategies addressing the unique concerns of different communities and compelling intrinsic motivations are more effective. This research offers valuable insights that can guide policymakers and healthcare workers in future emergencies to effectively communicate with the public by developing targeted messaging approaches to combat misinformation and educate the people.


\bibliography{references}


\newpage

\newpage
\appendix

\section{Datasets} \label{sec:datasets}

\subsection{Explanation of Datasets Used} \label{sec:explanationofdatasetsused}
We have collected, combined, and analyzed data from different sources. We created a comprehensive list in Table \ref{tab:comprehensive-dataset} to explain the sources and purpose of the datasets. The items in bold represent the datasets used in the models in this paper.

\begin{table}[!h]
  \centering
  \tiny
  \caption{Comprehensive list of datasets used and their purposes}
  \label{tab:comprehensive-dataset}
  \begin{tabular}{ rp{1.2in}p{1in}p{2.5in} }
    \toprule
    Sl. & Dataset & Source & Purpose \\
    \midrule
        1. & Completed Twitter Dataset & \citet{Chen2020-am} & A list of tweet IDs related to COVID-19. We collected the detailed tweets from this dataset as explained in Section \ref{sec:twitterdataset}. We collected \(15,768,845\) tweets to create this dataset \\
    \hline
        \textbf{2.} & \textbf{Twitter Dataset} & \textbf{Subset of the above} & \textbf{We created a stratified subset of \(466, 335\) tweets from the above dataset and used it for the experiments in this paper.} \\
    \hline
        3. & COVID-19 Rumor Dataset & \cite{Cheng2021-dc} & A labeled dataset containing COVID-19 rumors from news sources and Twitter, as explained in Section \ref{sec:covid19rumordataset}. \\
    \hline
        4. & Avax Tweets Dataset & \cite{Muric2021-fn} & A list of tweet IDs exhibiting antivaccine stance. We created a stratified sample of \(79,093\) detailed tweets from \(1.8\) million tweets as described in Section \ref{sec:avaxtweetsdataset}. \\
    \hline
        5. & Authentic Vaccination Tweets Dataset & \cite{Brandwatch2020-wh} & We manually collected vaccine-related tweets from trusted sources using Brandwatch as described in Section \ref{sec:authenticvaccinationtweetsdataset} \\
    \hline
        \textbf{6.} & \textbf{Motivation Training dataset} & \textbf{A combination of \(3\), \(4\), and \(5\) above} & \textbf{Combination of three sources to build the ground truth dataset for the training of de/motivating tweets classifier. The process is explained in Section \ref{sec:motivationtrainingdataset}.} \\
    \hline
        7. & Stance Tweets from Poddar & \cite{Poddar2021-xq} & A set of labeled tweets expressing the stance towards COVID-19 vaccination. The dataset is explained in Section \ref{sec:stancegroundtruthdataset} \\
    \hline
        8. & Stance Tweets from Cotfas & \cite{Cotfas2021-uj} & Another tweets dataset with label for stance towards COVID-19 vaccination, as explained in Section \ref{sec:stancegroundtruthdataset}. \\
    \hline
        9. & Manually Labeled Stance Dataset & Manually labeled & We collected a stratified sample of \(1,064\) tweets and manually labeled for stance towards COVID-19 vaccination. Details of the manual labeling are available in Section \ref{sec:stancegroundtruthdataset}. \\
    \hline
        \textbf{10.} & \textbf{Stance Ground Truth Dataset} & \textbf{A combination of \(7\), \(8\), and \(9\) above} & \textbf{We combined the data from three sources above for training the stance classifier for this paper. This dataset is explained in Section \ref{sec:stancegroundtruthdataset}.} \\
  \bottomrule
\end{tabular}
\end{table}

\subsection{Labeled Twitter Dataset} \label{sec:labeledtwitterdataset}

The dataset contains Tweet IDs along with the location and tweet timestamp. The tweets are labeled based on motivating/demotivating status, stance towards the COVID-19 vaccine, and topic in the tweet text. We removed the tweet texts and author information to comply with Twitter guidelines. You can use Hydrator API \citep{Documenting-the-Now2020-iy} to hydrate the tweets.

The anonymized dataset is available at:  \\
\href{https://zenodo.org/record/6842883/files/anonymized_tweets_with_labels.csv}{https://zenodo.org/record/6842883/files/anonymized\_tweets\_with\_labels.csv}

\section{Machine Learning Models} \label{sec:machinelearningmodels}

\subsection{De/Motivation Classifier} \label{sec:demotivationclassifier}

A pre-trained text classifier model to label the tweet texts as motivating or demotivating.

The model is available at:  \\
\href{https://zenodo.org/record/6842883/files/demotivation-classifier-distilbart-model.7z}{https://zenodo.org/record/6842883/files/demotivation-classifier-\\distilbart-model.7z}

\subsection{Vaccine Stance Classifier} \label{sec:vaccinestanceclassifier}

A pre-trained text classifier model to identify the stance of a tweet text towards the COVID-19 vaccine.

The model is available at:  \\
\href{https://zenodo.org/record/6842883/files/stance-detection-ct-BERT.7z}{https://zenodo.org/record/6842883/files/stance-detection-ct-\\BERT.7z}

\subsection{Topic Model} \label{sec:topicmodel}

A pre-trained topic model based on BERTopic to identify the topics in tweet texts.

The model for demotivating tweets is available at:  \\
\href{https://zenodo.org/record/6842883/files/topic-model-demotivating-bertopic.7z}{https://zenodo.org/record/6842883/files/topic-model-demotivating-\\bertopic.7z}

The model for motivating tweets is available at:  \\
\href{https://zenodo.org/record/6842883/files/topic-model-motivating-bertopic.7z}{https://zenodo.org/record/6842883/files/topic-model-motivating-\\bertopic.7z}

\section{Parameters for Reproducability} \label{sec:parametersforreproducability}

\subsection{Hardware Specification} \label{sec:reproducabilityhardware}
\begin{itemize}
    \item The data collection process was performed on a desktop computer with a Windows 10 setup containing 16GB RAM.
    \item The models were trained using a Google Colab Pro+ High-RAM (52GB) setup with an A100 GPU.
\end{itemize}

\subsection{De/Motivation Classifier} \label{sec:reproducabilitydemotivation}
Parameters for training the de/motivation classifiers are provided in Table \ref{tab:reproducability-demotivating}.

\begin{table}[!h]
  \centering
  \tiny
  \caption{Hyper-parameters for training the de/motivating classifiers}
  \label{tab:reproducability-demotivating}
  \begin{tabular}{ p{2.5in}rrr }
    \toprule
    Model & Batch Size & Epochs & Learning Rate\\
    \midrule
        DistilBERT (distilbert-base-uncased-finetuned-sst-2-english) & \(16\) & \(3\) & \(5e-5\) \\
    \hline
        RoBERTa (roberta-base) & \(16\) & \(3\) & \(1e-3\) \\
  \bottomrule
\end{tabular}
\end{table}

\subsection{Topic Model} \label{sec:reproducabilitytopicmodel}
Parameters for the topic modeling using BERTopic is provided in Table \ref{tab:reproducability-topic}.

\begin{table}[!h]
  \centering
  \tiny
  \caption{Parameters for re-training the BERTopic model}
  \label{tab:reproducability-topic}
  \begin{tabular}{ lr }
    \toprule
    Parameter & Value\\
    \midrule
        BERTopic version & \(0.9.4\) \\
    \hline
        Language & English \\
    \hline
        Calculate Probabilities & True \\
    \hline
        Verbose & True \\
  \bottomrule
\end{tabular}
\end{table}

\subsection{Vaccine Stance Classifier} \label{sec:reproducabilitystance}
Parameters for training the stance classifier is provided in Table \ref{tab:reproducability-stance}.

\begin{table}[!h]
  \centering
  \tiny
  \caption{Hyper-parameters for training the stance classifiers}
  \label{tab:reproducability-stance}
  \begin{tabular}{ p{2.5in}rrr }
    \toprule
    Model & Batch Size & Epochs\\
    \midrule
        RoBERTa (roberta-base) & \(16\) & \(10\) \\
    \midrule
        CT-BERT (digitalepidemiologylab/covid-twitter-bert-v2) & \(8\) & \(10\) \\
  \bottomrule
\end{tabular}
\end{table}

\section{Miscellaneous} \label{sec:miscellaneous}

\subsection{Topics from De/Motivating Classes} \label{sec:topicsfromdifferentclasses}

\subsubsection{Motivating Topics} \label{sec:appendixmotivatingtopics}
The top 10 frequent topics from the motivating tweets are displayed in table \ref{tab:motivating-topics} with the number of tweets in that topic and the top 5 words in that topic. 

\begin{table}[!ht]
\centering
\tiny
  \caption{Topics from motivating tweets}
  \label{tab:motivating-topics}
  \begin{tabular}{ p{0.3\columnwidth}lp{0.4\columnwidth} }
    \toprule
        Topic & Tweets & Top 5 words\\
    \midrule
        Face mask & 1452 & \emph{mask, masks, wear, wearing, face}\\
    \hline
        School & 1290 & \emph{schools, students, school, learning, teachers}\\
    \hline
        Optimism & 684 & \emph{amp, ov, optimistic, myself, ensure}\\
    \hline
        Sports & 480 & \emph{football, game, players, basketball, games}\\
    \hline
        {COVID-19} infection & 479 & \emph{sarscov2, sars, variant, infection, genome}\\
    \hline
        Biden inaugural & 435 & \emph{biden, joe, bidens, inaugural, transition}\\
    \hline
        Music & 431 & \emph{music, album, song, spotify, songs}\\
    \hline
        Voting & 423 & \emph{mail, voting, ballots, vote, mailin}\\
    \hline
        Death statistics & 386 & \emph{died, deaths, excess, 1000, per}\\
    \hline
        Cursing COVID-19 & 351 & \emph{corona, f*ck, sh*t, beer, b*tch}\\
    \bottomrule
  \end{tabular}
\end{table}

\subsubsection{Motivating Topics Related to ``Vaccine"}\label{sec:motivatingtopicvaccine}
Table \ref{tab:motivating-vaccine-topics} shows the top 10 ``vaccine" related topics from the motivating tweets along with the similarity score to the keyword ``vaccine."

\begin{table}[!hb]
\centering
\tiny
  \caption{Vaccine topics from motivating tweets}
  \label{tab:motivating-vaccine-topics}
  \begin{tabular}{ p{0.3\columnwidth}lp{0.4\columnwidth} }
    \toprule
        Topic & Similarity & Top 5 words\\
    \midrule
        Vaccine rollout & 0.77255 & \emph{desperately, vaccines, therapeutics, rollout, development}\\
    \hline
        Vaccination speed & 0.64514 & \emph{vaccineto, quicklyi, vaccinebut, bricker, coopting}\\
    \hline
        Flu during COVID-19 & 0.61227 & \emph{shot, flu, never, fightflu, twindemic}\\
    \hline    
        Misinformation and Trump & 0.60793 & \emph{misinformation, trust, pollquestion, trumpsvaccineisalie, hash}\\
    \hline    
        Infection prevention & 0.60710 & \emph{slim, iscovid19, contraceptive, prevent, immune}\\
    \hline    
        Vaccination directive & 0.60689 & \emph{vaccinationstha, adults, four, received, leading}\\
    \hline    
        Herd immunity & 0.58614 & \emph{herd, immunity, natural, seroprevalence, tcell}\\
    \hline    
        Flu season & 0.57635 & \emph{flu, influenza, cold, season, eradicated}\\
    \hline    
        Re-infection & 0.55869 & \emph{reinfection, suggesting, immunity, peop, hopef}\\
    \hline    
        Promoting vaccination & 0.54879 & \emph{vaccinat, protecting, stated, based, fully}\\
    \bottomrule
  \end{tabular}
\end{table}

\subsubsection{Demotivating Topics}\label{sec:appendixdemotivatingtopics}
The top 10 frequent topics from the demotivating tweets are displayed in table \ref{tab:demotivating-topics} with the number of tweets in that topic and the top 5 words in that topic. 

\begin{table}[!h]
\centering
\tiny
  \caption{Topics from demotivating tweets}
  \label{tab:demotivating-topics}
  \begin{tabular}{ p{0.3\columnwidth}lp{0.4\columnwidth} }
    \toprule
        Topic & Tweets & Top 5 words\\
    \midrule
        Nevada governor & 2723 & \emph{amp, wheelchairuser, haventwith, readthere, donaldyou}\\
    \hline
        Vaccine mandate & 1180 & \emph{vaccinated, vaccine, vaccines, stacontrolling, rollout}\\
    \hline
        New York governor & 1151 & \emph{cuomo, cuomos, andrew, nursing, homes}\\
    \hline    
        NFL COVID-19 restrictions & 871 & \emph{nfl, reservecovid19, rodgers, browns, aaron}\\
    \hline    
        COVID-19 in nursing homes & 801 & \emph{nursing, homes, patients, elderly, governors}\\
    \hline    
        COVID-19 restrictions for churches & 792 & \emph{church, pastor, churches, worship, easter}\\
    \hline    
        Cruise ship COVID-19 outbreak & 770 & \emph{cruise, ship, princess, navy, passengers}\\
    \hline    
        Promoting vaccination & 763 & \emph{vaccinated, vaccine, vaccines, minimizes, downleadership}\\
    \hline    
        Social distancing & 738 & \emph{distancing, social, practicing, yip, appa}\\
    \hline    
        Protective guidance & 709 & \emph{corona, virusmeaning, enveloped, asteroid, dieme}\\
    \bottomrule
  \end{tabular}
\end{table}

\subsubsection{Demotivating Topics Related to ``Vaccine"}
Table \ref{tab:demotivating-vaccine-topics} shows the top 10 ``vaccine" related topics from the demotivating tweets along with the similarity score to the keyword ``vaccine."

\begin{table}[!h]
\centering
\tiny
  \caption{Vaccine topics from demotivating tweets}
  \label{tab:demotivating-vaccine-topics}
  \begin{tabular}{ p{0.3\columnwidth}lp{0.4\columnwidth} }
    \toprule
        Topic & Similarity & Top 5 words\\
    \midrule
        Vaccine mandate & 0.90244 & \emph{vaccinated, vaccine, vaccines, stacontrolling, rollout}\\
    \hline
        Benefits of vaccination & 0.87415 & \emph{vaccinated, vaccine, vaccines, minimizes, downleadership}\\
    \hline
        Side effects & 0.77769 & \emph{covidvaccine, vaccinesideeffects, vaccinepasspos, fml, vaccinessavelives}\\
    \hline
        Immunity from vaccination & 0.66614 & \emph{immunity, natural, vaccinat, provides, protection}\\
    \hline
        Immunity and virus variants & 0.65232 & \emph{immune, immunity, autoimmune, newslasting, autoummune}\\
    \hline
        Promote vaccination & 0.63225 & \emph{getvaccinated, getvaccinatednow, resurrect, pandemichelp, getvaccinatedasapthe}\\
    \hline
        Immunocompromised & 0.58262 & \emph{immunocompromised, sacrifices, immunodeficient, guidelinescdc, judgment}\\
    \hline
        Vaccination statistics & 0.58117 & \emph{unvaccinated, 992, 995, peopleunvac, foundunvaccinated}\\
    \hline
        Natural immunity & 0.57885 & \emph{immunity, natural, itthat, isbig, covidnatural}\\
    \hline
        Vaccination in black community & 0.57702 & \emph{vaccinesaid, gonna, blacktwitter, cigarette, yall}\\
    \bottomrule
  \end{tabular}
\end{table}

\subsection{Sample Tweets from Different Classes} \label{sec:sampletweetsfromdifferentclasses}
We have listed a few sample tweets from different de/motivating classes and vaccination stances in Table \ref{tab:sample-tweets}. In this table, we presented the tweets in their original form without any data processing or cleanup for a better understanding of the text. We only removed any icons (emoticons) present in the text.

\begin{table}[!h]
  \centering
  \tiny
  \caption{Sample tweets from different de/motivating classes and stances}
  \label{tab:sample-tweets}
  \begin{tabular}{ p{3.5in}ll }
    \toprule
    Tweet & De/Motivating & Vaccination Stance \\
    \midrule
        Can stem cells treat COVID-19? Preliminary pre-clinical results show that lung-specific stem cells significantly reduce inflammation and lung tissue damage. & Motivating & Favor \\
    \hline
        Two of the last four years on \#Thanksgiving and all the other holidays, I was one of them. I’m proud of the work my coworkers and I did, all day, everyday (and night). \\
        For god’s sake, thank service industry workers this year and do NOT give them \#COVID19. & Motivating & Favor \\
    \hline
        BEST. VIDEO. ALL. YEAR. Please share with friends how the mRNA vaccine works to fight the coronavirus.\\
        NOTA BENE—The mRNA never interacts with your DNA. \#vaccinate \\
        (Special thanks to the Vaccine Makers Project @vaccinemakers of @ChildrensPhila). \#COVID19 & Demotivating & Favor \\
    \hline
        Yes, things are broken but they don't and won't change if people don't fight for it. \\
        Anyway, be safe and keep others safe. The only reason we still have covid is because of the selfish fucks going out like nothing's changed. & Demotivating & Favor \\
    \hline
        The policies peddled by incompetent health bureaucrats, weak politicians \& the group-thinking media of LOCKDOWN, MASK \& INJECT - all while suppressing Ivermectin is criminally negligent in its gross stupidity \& total insanity. & Demotivating & Against \\
    \hline
        For the Record, \\
        I am immune from ALL Mandatory Vaccinations. Mandatory vaccinations violate my Religious and Constitutional Rights and I do not consent to vaccinations of any kind, including COVID-19, the ``flu" and any/all other applicable ``diseases." & Demotivating & Against \\
    \hline
        Re-infection with COVID is extremely rare. Our immune systems work remarkably well against SARS-Cov-2. Anyone saying otherwise is uninformed (not reading the studies/not seeing covid patients) or lying. & Motivating & Against \\
    \hline
        Not all 35 million took the flu vaccine. A lot less would catch it if they took it, but the danger of dying with the flu is much less than with Covid-19, so many don't bother. Once there is a safe vaccine for C19, I hope many, many more people take it than take a flu vaccine. & Motivating & Against \\
  \bottomrule
\end{tabular}
\end{table}

\end{document}